\documentclass[jcp,onecolumn,floatfix]{revtex4-1}

\usepackage{graphics} 
\usepackage{graphicx} 
\usepackage{color} 
\usepackage{bm} 
\usepackage[dvipsnames]{xcolor,colortbl}
\usepackage{amsmath}
\usepackage{amssymb}
\usepackage{amsfonts}

\usepackage[normalem]{ulem}  

\usepackage{xspace}
\newcommand{\ie}{\textit{i.e.}\xspace}

\newcommand{\I}{\textrm{I}}
\newcommand{\II}{\textrm{I\kern -0.3ex I}}
\newcommand{\III}{\textrm{I\kern -0.3ex I\kern -0.3ex I}}
\newcommand{\IV}{\textrm{I\kern -0.3ex V}}
\newcommand{\Vs}{\textrm{V}}
\newcommand{\Ix}{$\mathrm{I}_x$\xspace}
\newcommand{\IIx}{$\mathrm{II}_x$\xspace}
\newcommand{\Honey}{$\mathrm{H}$\xspace}
\newcommand{\SnubSquare}{$\mathrm{S}_1$\xspace}
\newcommand{\SnubSquareTwo}{$\mathrm{S}_2$\xspace}
\newcommand{\Pone}{$\mathrm{P}_1$\xspace}
\newcommand{\Ptwo}{$\mathrm{P}_2$\xspace}
\newcommand{\Pthree}{$\mathrm{P}_3$\xspace}
\newcommand{\Vx}{$\mathrm{V}_x$\xspace} 

\newcommand{\Genome}                 {\mathcal{X}}
\newcommand{\kmeans}[0]{$\mathbf{k}^{*}_{32}$-clustering\xspace}
\newcommand{\kstar}[1]{$\mathbf{k}^{*#1}_{32}$\xspace}

\begin{document} 
\title{Ordered ground state configurations of the asymmetric Wigner bilayer system -- revisited: an unsupervised clustering algorithm analysis}

\author{Benedikt Hartl$^{1}$}
\author{Marek Mihalkovi{\v c}$^{2}$}
\author{Ladislav \v{S}amaj$^{2}$} 
\author{Martial Mazars$^{3}$} 
\author{Emmanuel Trizac$^{3}$} 
\author{Gerhard Kahl$^{1}$} 

\affiliation{$^1$Institute for Theoretical Physics and Center for
  Computational Materials Science (CMS), TU Wien, Austria
  \\$^2${Institute of Physics, Slovak Academy of Sciences, Bratislava,
  Slovakia} \\ $^3${Universit{\'e}  Paris-Saclay, CNRS, LPTMS, Orsay, France}}


\begin{abstract}
We have re-analysed the rich plethora of ground state configurations of the asymmetric Wigner bilayer system that we had recently published in a related diagram of states [M. Antlanger \textit{et al.}, Phys. Rev. Lett. \textbf{117}, 118002 (2016)], comprising roughly 60~000 state points in the phase space spanned by the distance between the plates and the charge asymmetry parameter of the system. In contrast to this preceding contribution where the classification of the emerging structures was carried out ``by hand'', we have used this time machine learning concepts, notably based on a principal component analysis and a $k$-means clustering approach: using a 30-dimensional feature vector for each emerging structure (containing relevant information, such as the composition of the configuration as well as the most relevant order parameters) we were able to re-analyse these ground state configurations in a considerably more systematic and comprehensive manner than we could possibly do in the previously published classification scheme. Indeed we were now able to identify new structures in previously unclassified regions of the parameter space and could considerably refine the previous classification scheme, identifying thereby a rich wealth of new emerging ground state configurations. Thorough consistency checks confirm the validity of the newly defined diagram of states. 
\end{abstract}

\date{\today}

\maketitle


\section{Introduction}
\label{sec:introduction}

Scientists nowadays are often confronted with huge data sets, may it be images or written text of any kind, scattering data from particle detectors, geometric data of lattice structures (\ie, particle arrangements) generated by experiments, theoretical frameworks, or via computer simulations~\cite{HIGHBIAS2019}. Analyzing such data sets is usually a task far from being trivial, especially in high-dimensional spaces: let a data set, $\mathbf{X}=\{\mathbf{x}_1,\dotsc\mathbf{x}_N\}$, consists of $N$ data elements (or, equivalently, data points), termed $\mathbf{x}_i$ (with $i = 1, \dotsc, N$); then each data element can be viewed as a vector, $\mathbf{x}_i = \{x_1,\dotsc,x_{N_f}\}$ (also referred to as feature vector) which contains a large number of so-called features, $x_j$ (with $j = 1, \dotsc, N_f$); examples for such features are the values of different pixels of an image, the different channels of the measurement of a particle collision experiment or the coordinates, orientations and/or order parameters of a particle arrangement of a (lattice) structure. Under certain circumstances a few clear signals in the data (\ie, a few characteristic features, $x_j$, in the feature vectors, $\mathbf{x}_i$, of a data set $\mathbf{X}$) may allow us to categorize the elements of a data set, for instance into different structural families with well-defined, characteristic order parameters. However, the shear amount of typical data sets and the often immense complexity of the involved features usually render a classification scheme intractable to be manually carried out by a human being. In such cases and in an effort to obtain a more comprehensive picture of the properties of the underlying data in general it is highly advisable to use methods from unsupervised machine learning~\cite{HIGHBIAS2019, RokachandMaimon2010}.
For example, an approach based on neural network potentials was used for local structure detection in polymorphic systems \cite{Geiger2013} and dimensionality reduction techniques were successfully utilized for crystal classification \cite{vanDamme2020}.
Very recently, an unsupervised topological learning approach was proposed for identifying atomic structures \cite{Becker2022} and crystal nucleation \cite{Becker2022a} without \textit{a priori} knowledge of the underlying physical system.

In this contribution, we reconsider and reanalyze the diagram of ground state configurations (occurring at vanishing temperature) of the so-called asymmetric Wigner bilayer system. In such a system point charges form on each of the oppositely charged, confining planar walls (${\cal L}_1$ and ${\cal L}_2$) ordered particle configurations; the respective surface charges of the plates, $\sigma_1$ and $\sigma_2$, which are separated by a reduced, dimensionless distance $\eta$ are not necessarily identical -- hence the system is termed asymmetric; the ratio of the surface charge densities is defined as $A = \sigma_2/\sigma_1$ (with $A \in [0, 1]$). The entire system is charge neutral. Depending on the location in the parameter space (spanned by $A$ and $\eta$) the system assumes ground state configurations that are characterized by the composition of the system $x = N_2/N$ (with $N_2$ being the number of particles occupying ${\cal L}_2$). Thus for a given pair  $(A, \eta)$ the ground state configuration is specified by the value of $x$ and by the lattices formed on the two layers (e.g., in terms of the respective lattice vectors). In preceding contributions \cite{Antlanger2014, MoritzAntlanger2015, Antlanger2016, Antlanger2018} these ground state configurations were determined via suitably adapted optimization tools (based notably on evolutionary algorithms), limiting -- for numerical reasons -- the number of particles per unit cell to $N = 40$. In a subsequent step the emerging ground state configurations were classified in terms of the emerging structures, based on suitably defined order parameters \cite{Steinhardt1983, Lechner2008}. This classification procedure, which was realized by ``hand'' has led to a highly intricate diagram of states, where in total 14 different structures could be identified.

This ``manual'' scheme is of course prone to fail when aiming at an exhaustive classification of the emerging structures. To overcome this drawback we have re-analyzed in this contribution the available set of data using tools of machine learning; to be more specific we have employed a clustering algorithm technique from unsupervised learning to classify these data in terms of clusters of data: in this way, we collect data elements into distinct subgroups that share similar (structural) features. We start the related analysis by defining a suitable feature vector (with length $N_{\rm f}$) which captures for each specific lattice structure the relevant information: in our case $N_{\rm f} = 30$ and the feature vector contains as elements the composition, a selected choice of order parameters, and some information about the radial distribution function \footnote{At this point it should be mentioned that also other related techniques are available which may even directly operate on the structural data (\ie, on a data set of coordinates of the particles) -- see for instance Refs.~\cite{Jadrich2018, Jadrich2018a}.}. These feature vectors are vectors in the $N_{\rm f}$-dimensional feature space, spanned by the $N_{\rm f}$ above-mentioned features. The aim of this procedure is to identify spatially separated clusters of similar data elements in this space: in this sense one specific cluster is considered to contain similar (structural) data elements \footnote{See Ref.~\cite{RokachandMaimon2010} for an in-depth discussion on different, problem specific similarity measures in data science problems. Methods from unsupervised machine learning \cite{HIGHBIAS2019} can be used to analyze a data set of feature vectors (or of order parameters in our case) for certain similarity measures in the features which may permit us to algorithmically organize the elements of the data set into an initially unknown set of categories \cite{RokachandMaimon2010}.}. 
To be more specific we start off with the feature vectors but reduce in a subsequent step the complexity of the problem via a so-called principal component analysis which maps (in the so-called dimensional reduction step) the huge amount of data into a low-dimensional latent space representation capturing and maintaining/preserving the relevant aspects (or features) of each data element and discarding the other, rather irrelevant information. Based on this reduced information we then start to sort the different structures into clusters. In this contribution we use a rather simple (and therefore presumably the most applied) form of unsupervised machine learning \cite{HIGHBIAS2019}, namely \textit{clustering algorithms}~\cite{Moghadam2019, Boattini2019, HIGHBIAS2019}, in order to organize data sets of lattice structures into families of structures. In the language of clustering algorithms, the procedure of categorizing data elements via a suitably defined similarity measure between data points in the feature space into different clusters is usually denoted as \textit{clustering} or \textit{labeling}: each of the $N$ elements of a data set is labeled by an identifier, $k_c$, which assigns each of its element to one of the categories (or \textit{clusters}) identified by the clustering algorithm \footnote{Depending on the particularly applied clustering algorithm the number of clusters, $N_c$, maybe a preset parameter to the algorithm or may even be identified by the algorithm during execution.}.

The size, $N$, of the data set and the dimensionality, $N_f$, of the feature space may influence the choice of clustering algorithm used in practice (from a conceptual point of view).

The data set size and feature space dimensionality are also relevant when considering the time complexity of different clustering algorithms (with respect to $N$ and $N_f$) and the available resource when applying a specific method.

In a first step we have applied the above-outlined two-step procedure to the case of the {\it symmetric} Wigner bilayer system, where the charge densities on ${\cal L}_1$ and ${\cal L}_2$ are equal. We find that the principal component analysis indicates that only a five-dimensional latent space is required. We recover -- not surprisingly, as anticipated by the exact results \cite{Samaj2012, Samaj2012a} -- that the emerging structures can be classified into the five well-known clusters, each representing one of the well-known ordered ground state structures. With this confirmation of our procedure in mind, we proceed to the {\it asymmetric} Wigner bilayer system, where the aforementioned ``by hand'' classification \cite{MoritzAntlanger2015, Antlanger2016, Antlanger2018} has led to 14 structural classes. Using the same 30-dimensional feature vector the principal component analysis provides evidence that the feature space can be mapped into a nine-dimensional latent space. Based on this reduced representation we then perform the $k$-means clustering step of the structural data. Eventually and performing numerous consistency checks of this classification scheme we end up with a reliable classification of the structural data into 32 clusters which (i) identifies new, so far unclassified structures and (ii) does not leave any white regions in the diagram of states.  
We thereby demonstrate that our unbiased classification scheme is considerably more reliable than a biased search ``by hand''.

The manuscript is organized as follows: in the subsequent section, we briefly summarize the essential features of the asymmetric Wigner bilayer system and outline how the energy of the ordered configurations can be evaluated with high numerical accuracy. Further, we define in this Section the order parameters that we have used to characterize the emerging structures; we introduce our machine learning-based methods of how to identify structural similarities in our unlabeled set of data of ordered structures: the principal component analysis and the $k$-means algorithm.  In Section III the discussion of the results starts with a brief discussion of the originally derived diagram of states and the specific steps of how the unsupervised clustering algorithm is applied to the Wigner bilayer system. We then discuss the emerging results with the previously known exact results of the {\it symmetric} Wigner bilayer and then proceed to the {\it asymmetric} case, where particular focus is laid on the emerging new insights. The body of the manuscript is closed with a Conclusion. Three Appendices close the manuscript, they are dedicated to conceptual details as well as to in-depth discussions of particular features of the machine learning-based approaches and provide more detailed background information.



\section{Model and methods}
\label{sec:models_methods}

\subsection{Model} 
\label{subsec:model}

In the Wigner bilayer system classical, negative point charges $q = - e$ ($e$ being the elementary charge) are confined between two parallel plates (${\cal L}_1$ and ${\cal L}_2$) that are separated by a distance $d$. The plates carry uniform, positive surface charge densities, $e\,\sigma_1$, and $e\,\sigma_2$, which are not necessarily equal. The total system is electro-neutral. 

\begin{figure}[htbp]
\begin{center}
\includegraphics[width=8cm]{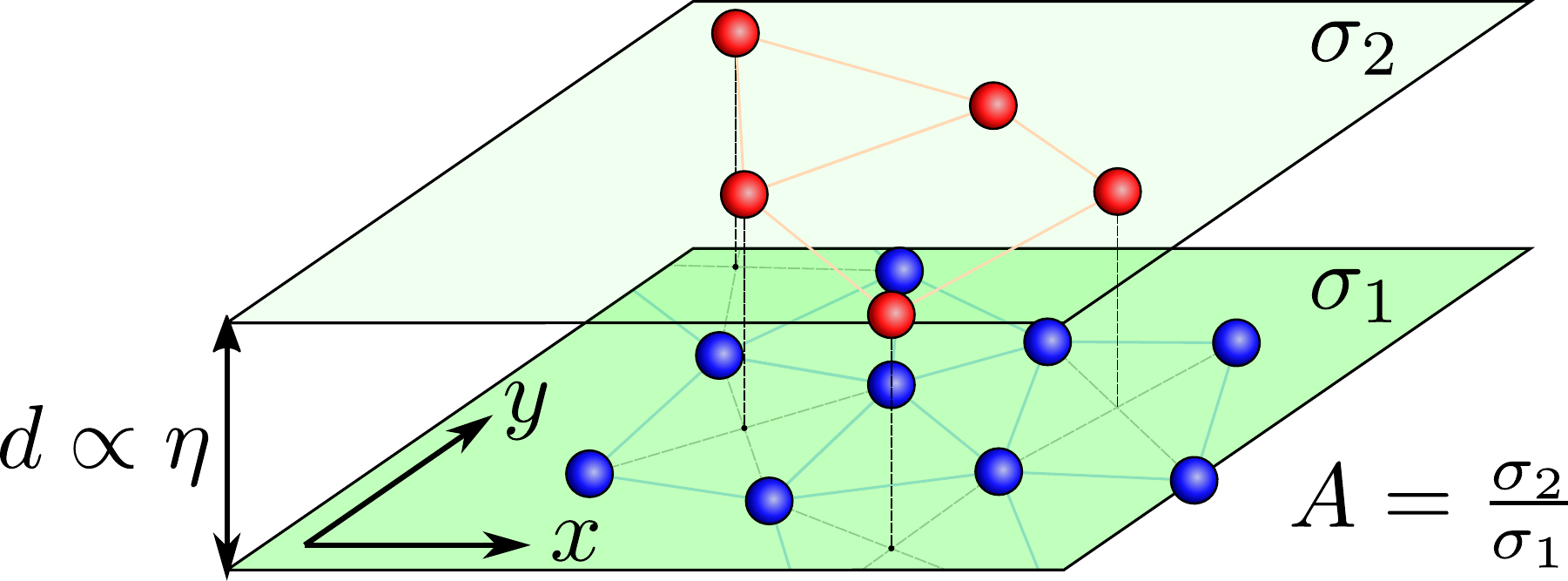}
\caption{Schematic view of the classical Wigner bilayer system: confined between two parallel plates, which are separated in the $z$-direction by the distance $d$, classical point charges (colored in blue and red) form ordered configurations on these plates (carrying homogeneous surface charge densities $e\,\sigma_1$ and $e\,\sigma_2$). The $x$- and $y$-directions of the Cartesian coordinate system are indicated.}
\label{fig:model}
\end{center}
\end{figure}

Being interested in the ground state configurations of the system, the energetically most stable configurations that the particles form at vanishing temperature $T$, we can rely on Earnshaw's theorem \cite{Earnshaw1848} which states that the particles have to be located on the plates. For a schematic view of the setup, we refer to Fig.~\ref{fig:model}. 

The system is thus characterized by two parameters: 

\begin{itemize}
\item [(i)] the ratio of the surface charge densities, A, 

$$
A = \frac{\sigma_2}{\sigma_1}
$$
which -- without loss of generality -- can be assumed to lie within the interval $[0, 1]$; a Wigner bilayer system with $A \equiv 1$ is termed symmetric, while otherwise it is called asymmetric;
\item[(ii)] the distance between the plates, $d$, which -- for convenience -- is reformulated via a dimensionless parameter $\eta$, defined as

$$
\eta = d \sqrt{\frac{\sigma_1 + \sigma_2}{2}}.
$$
\end{itemize}
Assuming that $N$ particles populate the unit cell, we denote by $N_1$ and $N_2$ the number of particles that populate ${\cal L}_1$ and ${\cal L}_2$, respectively; obviously $N = N_1 + N_2$ and we define the composition of the system $x$ via $x = N_2/N$. The area particle number density is denoted by $\rho$, which we set -- without loss of generality -- to unity.

\subsection{Energy calculations}
\label{subsec:lattice_sum}

In the classical Wigner bilayer system, the point charges interact via the long-range Coulomb interaction with each other and with the uniformly charged plates. Further, there is a distance-dependent, but otherwise, constant plate-to-plate interaction contributing to the total (internal) electrostatic energy of the unit cell of a bilayer structure, $E(\mathbf r^{N};A,\eta)$, which is given in Gauss units by \cite{MoritzAntlanger2015, Antlanger2016, Antlanger2018}

\begin{equation}
    E(\mathbf r^{N};A,\eta) = \sum_{i=1}^N \left[
        \sum_{j=1}^N \sum_{\mathbf{S}_\mathbf{n}}{\vphantom{\sum}}^* \frac{e^2}{|\mathbf r_i - \mathbf r_j + \mathbf{S}_\mathbf{n}|}
    - 2\pi e^2(\sigma_1-\sigma_2) z_i \right] + {\rm const.}
    \label{eq:results:wigner:energy:coulomb}
\end{equation}
$\mathbf r^N=(\mathbf r_1,\dotsc,\mathbf r_N)$ is the set of position vectors $\mathbf r_i=(x_i, y_i, z_i)$ of the $N$ point charges in the unit cell, $z_i=0$ or $z_i=d ~(\propto \eta)$ specifies if particle $i$ occupies ${\cal L}_1$ or ${\cal L}_2$; finally $\mathbf{S}_\mathbf{n}$ is a symbolic notation for periodic images of the unit cell in the $x$- and $y$-directions used to carry out the lattice summation \cite{Mazars2011}. In Eq.~(\ref{eq:results:wigner:energy:coulomb}) the symbol $\sum^*$ indicates, that for $\mathbf{S}_\mathbf{n}=(0,0,0)$ the sum is carried out only for $j>i$ to avoid double counting within the unit cell (see Refs. \cite{MoritzAntlanger2015, Antlanger2016, Antlanger2018} for details). For convenience, we chose the dielectric constant, $\epsilon$, of the medium into which the particles are immersed, as well as the dielectric constant of the two plates, $\epsilon_1$ and  $\epsilon_2$, to be equal; henceforward we set this value to unity, \ie, $\epsilon=\epsilon_1=\epsilon_2=1$. Following closely  Ref.~\cite{MoritzAntlanger2015} we employ Ewald summation techniques~\cite{Ewald1921} -- specifically implemented for quasi-2D bilayer geometries~\cite{MoritzAntlanger2015, Mazars2011} -- to numerically evaluate the long-range electrostatic energy of the system in a highly reliable and computationally efficient manner.

Searching for a given pair of $\eta$ and $A$ for the global ground state configuration of the asymmetric Wigner bilayer system boils down to identifying simultaneously the correct number of particles per unit cell ($N$), to finding the optimal arrangement of the particles on the two plates, $\mathbf{r}^N$, and to identifying the correct unit cell geometry; from the positions, $\mathbf r^N$, one can extract $N_1$ and $N_2$ and thus the composition $x$. The components $a_{11}, a_{21}$, and $a_{22}$ of the vectors $\mathbf{a}_1=(a_{11},0,0)$ and $\mathbf{a}_2=(a_{21},a_{22},0)$ that define the unit cell are subject to the structure optimization problem under the constraint of keeping the area of the unit cell, $S_0=a_{11}\,a_{22}$, constant; the lattice vector $\mathbf{a}_3=(0,0,d)$ is fixed by the plate separation distance $d$. With all this in mind, we minimize in our search for the ground state configuration the total energy per particle, $E(\mathbf r^{N};A,\eta)/N$, as specified in  Eq.~(\ref{eq:results:wigner:energy:coulomb}).

Henceforward we collect all these variational parameters via the following short-hand notation: 

\begin{equation}
    \Genome \equiv (\mathbf{r}^N, \mathbf{a}_{1}, \mathbf{a}_{2}) .
    \label{eq:results:wigner:config}
\end{equation}
In this spirit, we can also write $E(\mathbf{r}^N;A,\eta) \equiv E(\Genome;A,\eta)$ to parameterize the energy. If -- at one occasion or the other -- the particular values of $A$ and $\eta$ are not of relevance for the discussion we then simply write $E(\Genome)$ or even drop the argument of the energy completely, \ie, we simply use $E$.

The accuracy required for the evaluation of the energies of competing structures is tremendously high: anticipating that $E/\left(N\sqrt{\rho}e^2\right)\approx -1$, setting thus the energy of the system, we note that relative differences in the energies of competing structures down to the sevenths or eighths digit are quite frequent. In view of these accuracy requirements, the search for the ground state configurations becomes thus a very delicate optimization problem in a high dimensional search space; the first attempt to solve this challenge has been successfully carried out in Refs.~\cite{MoritzAntlanger2015, Antlanger2016, Antlanger2018} with the help of memetic evolutionary algorithms.

The computational cost for exploring the high-dimensional search space can be reduced since the energy, $E/N$, defined in Eq.~(\ref{eq:results:wigner:energy:coulomb}), can be split into a (i) structure-dependent, but $A$-independent contribution and into an (ii) $A$-dependent, but structure-independent contribution. Following Refs.~\cite{MoritzAntlanger2015, Antlanger2016, Antlanger2018} we first define  the reduced energy per particle as

\begin{equation}
    \frac{E^*(\mathbf r^N;A,\eta)}{N} \equiv \frac{E(\mathbf r^N;A,\eta)}{N\sqrt{\rho}e^2} .
    \label{eq:results:wigner:energy:unitcell}
\end{equation}
We then identify the structure-independent contribution to $E^*$ as

\begin{equation}
    \frac{E^*_A(A,\eta,x)}{N} = 2^{3/2}\pi\eta\frac{A}{(1+A)^2}[A-2x(1+A)],
    \label{eq:results:wigner:energy:trick}
\end{equation}
leading thus to 

\begin{equation}
    \frac{E^*(\mathbf r^N;A,\eta)}{N} = \frac{1}{N}\left[E^*(\mathbf r^N;A_0,\eta) -  E^*_A(A_0,\eta,x) + E^*_A(A,\eta,x)\right]
    \label{eq:results:wigner:energy:reduced}
\end{equation}
with the reference asymmetry parameter $A_0$ (which, without loss of generality, we set to $A_0=0$) and the composition $x$ introduced above. With this reformulation of $E^*$, the computational cost of the identification of ground state configurations can be substantially reduced but remains nevertheless quite high. 

Using the above separation of the internal energy, the ground state configurations have been identified in Ref.~\cite{MoritzAntlanger2015} via independent evolutionary searches at a fixed value of $A(=0)$ for different, numerically tractable values of the composition, $x$, on a fixed grid for the plate separation distance parameter $\eta~ (\geq 0)$.  The resulting set of structural ground state configurations obtained for different compositions, identified at $A=0$ but for a particular value of $\eta$, provide all necessary information to identify subsequently the ground state configuration for any state point $(\eta, A)$; in this manner, the approach becomes highly efficient. Limiting for computational reasons the number of particles per unit cell to $N = 40$, one can calculate \cite{Hartl2020Thesis} that at each value of $\eta$ the total number of possible compositions is $N_{\rm tot} = 401$ \footnote{The total number of possible compositions at each value of $\eta$ is $N_\mathrm{tot} = 1 + \frac{1}{2}  \sum_{n=2}^{N} (n - n \mod 2 )$ given that $0 < N_2 \leq \frac{N_1}{2}$ for $N_1>1$, and only counting the monolayer structure with $N_2=0$ and $N_1=1$ once. For $N=40$ we thus have $N_\mathrm{tot} = 401$.}. Further and following Ref.~\cite{MoritzAntlanger2015}, we chose in our numerical analysis a uniform grid of $N_{\eta}=141$ different values for $\eta\in[0, 1]$ (resulting in $N_{\rm tot}\times N_{\eta}=56541$ different evolutionary optimized structures in total), and specify a uniform grid of $N_A=201$ values for the asymmetry parameter $A\in[0, 1]$. The numerical values for the grid in $\eta$ and $A$ are thus $\Delta \eta=10^{-2}/\sqrt{2}$ and $\Delta A=5\times10^{-3}$. 

For given values of $\eta$ and $A$ the configuration which minimizes $E^*(\Genome)/N$ is considered as the related ground state configuration and we denote the ground state  energy as $E^*_\mathrm{GS}(A,\eta)/N$. Henceforward, we usually drop the arguments of the energy -- unless we want to emphasize its dependency on certain arguments -- and we synonymously use $E^*/N$ for the expression given in  Eq.~(\ref{eq:results:wigner:energy:reduced}) and $E^*_\mathrm{GS}/N$ for the ground state energy, respectively.

\subsection{Order parameters}
\label{subsec:order_parameters}

\begin{figure}[htbp]
\begin{center}
\includegraphics[width=6cm]{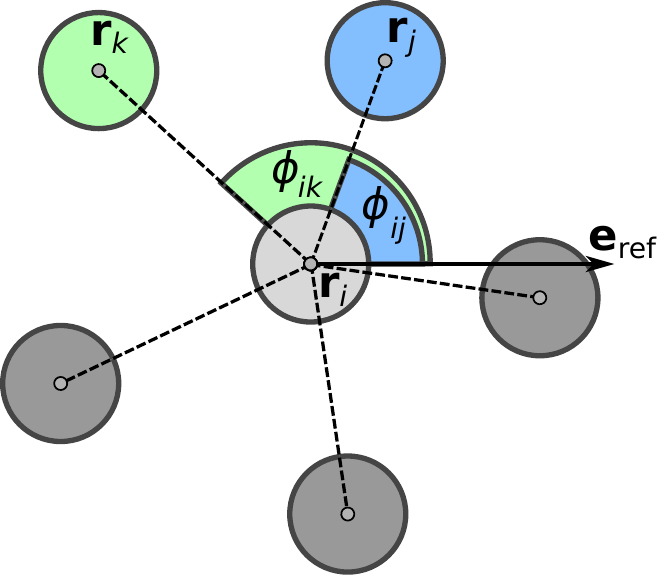}
\caption{Schematic representation of a tagged particle (with index $i$) and of its neighbouring particles (with indices $j$ and $k$), as defined in a preceding Voronoi construction. The angles $\phi_{ij}$ and $\phi_{ik}$ are enclosed by the related interparticle bonds and a reference unit vector, $\mathbf{\hat e}_\mathrm{ref}$.}
\label{fig:boop}
\end{center}
\end{figure}

The identification of the ordered ground state configurations was based in Refs.~\cite{MoritzAntlanger2015, Antlanger2016, Antlanger2018}. In this contribution, we specifically make use of the composition of the system, $x$, and on bond-orientational order parameters (BOOPs) $\Psi_\nu = \Psi_\nu(\Genome)$. In their most elementary version, these parameters are defined as

\begin{equation}
    \Psi_\nu(\Genome)=\frac{1}{N}\sum\limits_{i=1}^N\left|\frac{1}{\mathcal{N}_i} \sum\limits_{j=1}^{\mathcal{N}_i}\exp[\imath\nu\phi_{ij}] \right| .
    \label{eq:bond_orientational_order_parameter}
\end{equation}
For a tagged particle with index $i$ the angles $\phi_{ij}$ are enclosed by the bond of particle $i$ to one of its ${\cal N}_i$ neighbouring particles $j$ and some reference axis $\mathbf{\hat e}_\mathrm{ref}$ (see Fig.~\ref{fig:boop}); thus the angles $\phi_{ij}$ are given by $\cos\phi_{ij}=\mathbf{\hat r}_{ij}\cdot\mathbf{\hat e}_\mathrm{ref}$, with $\mathbf{\hat r}_{ij} = (\mathbf{r}_{j} - \mathbf{r}_{i})/|\mathbf{r}_{j} - \mathbf{r}_{i}|$. The neighbors are identified via a standard Voronoi construction \cite{LejeuneDirichlet1850, Voronoi1908} using an open-source software package \cite{Jones2001}. 

The orientational symmetry of the neighborhood of particle $i$ is characterized by the (integer) variable $\nu$: the $\nu$-fold BOOP $\Psi_\nu(\Genome)$ assumes the value one if the angles between neighbors are multiples of $2\pi/\nu$ and attain values close to zero for a disordered particle arrangement or if there is no $\nu$-fold symmetry.

Due to small, inherent numerical inaccuracies, the lattices that we deal with are never perfect; consequently, the exact number of nearest neighbors can be strongly influenced by minute changes in the particle positions, making the actual evaluation of the BOOPs numerically unstable. In an effort to guarantee better numerical stability in the evaluations of BOOPs we modify via a simple remedy the BOOPs specified in Eq.~(\ref{eq:bond_orientational_order_parameter}) by including a weight factor that is related  to the polygon side length, $l_{ij}$, that neighboring particles (with indices $i$ and $j$) share \cite{Mickel2013}
\begin{equation}
    \Psi_\nu(\Genome)=\frac{1}{N}\sum\limits_{i=1}^N
    \left|
    \frac{1}{L_i}
    \sum\limits_{j=1}^{\mathcal{N}_i}l_{ij}\exp[\imath\nu\phi_{ij}]
    \right|,
    \label{eq:bond_orientational_order_parameter:weighted}
\end{equation}
where $L_i=\sum_{j=1}^{\mathcal{N}_i} l_{ij}$; the $l_{ij}$ are again extracted from the Voronoi construction.

Henceforward we usually drop the argument, $\Genome$, and often simply write $\Psi_\nu\equiv\Psi_\nu(\Genome)$, unless the explicit indication of a particular realization of a structure, $\Genome$, is required for the discussion.

The above definition of the BOOPs has been extended by considering separately particles in different layers (or combinations thereof). Following Refs. \cite{MoritzAntlanger2015,Antlanger2016,Antlanger2018}
we use the following four variations of BOOPs:

\begin{itemize}
    \item [(1)] $\Psi^{(1)}_\nu$, quantifying $\nu$-fold bond-orientational order of particles in ${\cal L}_1$;
    \item [(2)] $\Psi^{(2)}_\nu$, quantifying $\nu$-fold bond-orientational order of particles in ${\cal L}_2$;
    \item [(3)] $\Psi^{(3)}_\nu$, quantifying $\nu$-fold bond-orientational order of particles of both layers projected onto the same plane;
    \item [(4)] $\Psi^{(4)}_\nu$, quantifying $\nu$-fold bond-orientational order of particles of layer two, considering only layer one particles (projected onto the same layer) as neighbors.
\end{itemize}

For convenience, we introduce here a short-hand notation for addressing a set of $u_i=u_1,\dotsc,u_n$ different BOOPs, $\Psi^{(u_i)}_{\nu_j}$, of different bond-orientational order, $\nu_j=\nu_1,\dotsc,\nu_m$, via
$\Psi^{(u_1, \dotsc, u_n)}_{[\nu_1, \dotsc, \nu_m]} = \{
    \Psi^{(u_1)}_{\nu_1},\dotsc,\Psi^{(u_1)}_{\nu_m},
    \Psi^{(u_2)}_{\nu_1},\dotsc,\Psi^{(u_2)}_{\nu_m},
    \dotsc,
    \Psi^{(u_n)}_{\nu_1},\dotsc,\Psi^{(u_n)}_{\nu_m}
\}$;
if only one lower index is used, e.g. $[\nu]$, the lower brackets may also be omitted and we may write $\Psi^{(u_1, \dotsc, u_n)}_{\nu}=\{\Psi^{(u_1)}_{\nu},\dotsc,\Psi^{(u_n)}_{\nu}\}$.


\subsection{Identifying similarities in unlabeled data: Clustering algorithms}
\label{subsec:unsupervised_clustering}

\subsubsection{General remarks}
\label{subsubsec:general_remarks}

In this contribution, we are ultimately interested to analyse systematically the ground state configurations of the asymmetric Wigner bilayer system in terms of families (or clusters) of similar (or related) structures, with the configurations being characterized notably via their composition and their order parameters. We emphasize that there is no prior information in our data set that relates the different data elements (\ie, the different structures) to certain crystalline phases, thus our task is to classify unlabeled data. In an effort to achieve the above goal, we have used clustering algorithms techniques from unsupervised learning that have received a rapidly growing share of interest in recent years. This interest is related to the urgent need to cope with ever-growing, huge sets of data (with often more than a million elements, as it occurs in our case), each of them being characterized by sometimes several thousands of so-called features, and to classify these data in terms of clusters of data (\ie, to group the data elements into distinct subgroups that share similar features). In the following, we briefly introduce and summarize two schemes: (i) the principal component analysis (PCA) and (ii) the $k$-means clustering -- that have been designed to identify and classify huge sets of data in terms of a small number of characteristic features.

To specify the notation we introduce a data set ${\bf X} = \{ {\bf x}_1, \dots, {\bf x}_N \}$ which consists of $i = 1, \dots, N$ data elements (or data points) ${\bf x}_i$. Each ${\bf x}_i$, also referred to as feature vector, may contain a large number ($N_{\rm f}$) of so-called features $x_i$, \ie, ${\bf x}_i = \{ x_1, \dots, x_{N_{\rm f}} \}$; the ${\bf x}_i$ are elements in the $N_{\rm f}$-dimensional, so-called feature space. With all this in mind, ${\bf X}$ can also be viewed as an ($N \times N_{\rm f}$)-matrix. The huge size of typical data sets stored in ${\bf X}$ and the often immense complexity of the involved features usually renders a conventional classification scheme of these data in terms of families or clusters of data intractable to be manually carried out by a human being. In such cases and to obtain a more comprehensive picture of the properties of the underlying data, it has turned out to be advantageous to involve concepts from unsupervised machine learning \cite{HIGHBIAS2019, RokachandMaimon2010}. 

For the problem addressed in this contribution, we use the possibly simplest and hence maybe the most frequently used form of unsupervised machine learning \cite{HIGHBIAS2019}, namely (unsupervised) clustering algorithms \cite{Moghadam2019, Boattini2019, HIGHBIAS2019}; as we will demonstrate in this contribution such an approach turns out to be very helpful to organize data sets of lattice structures of the classical Wigner bilayer system into families of structures  in an unsupervised manner, \ie, without prior knowledge about structure- or feature-dependent relations in our unlabeled data set. 

In the following we will introduce two basic concepts that help us to implement our approach: in the subsequent paragraph, we will introduce the so-called \textit{principal component analysis} which helps us to identify and capture characteristic features in a data set in an effort to reduce the dimensionality of the feature space. This reduction is done in view of the fact that operating in a high-dimensional feature space makes a meaningful cluster analysis very difficult. Instead, experience tells us that the relevant information is often captured by a few, characteristic features; therefore it is the dedicated aim of our approach to identifying these relevant features. In the subsequent paragraph, we present a frequently used standard clustering algorithm, namely the \textit{$k$-means clustering}.

\subsubsection{Principal component analysis (PCA)}
\label{subsubsec:PCA}

Working now in a high, $(N \times N_{\rm f})$-dimensional space of data we try to reduce the dimensionality of the $N_{\rm f}$-dimensional feature space to a considerably smaller, $N_\ell$-dimensional and so-called latent space, using the PCA. Our aim is thus to transform a data set ${\bf X} = \{ {\bf x}_1, \dots, {\bf x}_N \} \in \mathbb{R}^{N \times N_{\rm f}}$ into a low-dimensional latent space representation of the data, ${\bf L} = \mathbb{P}_{XL} ({\bf X}) = \{ {\bf l}_1, \dots, {\bf l}_N \} \in \mathbb{R}^{N \times N_{\rm \ell}}$; the ${\bf l}_i \in \mathbb{R}^{N_{\ell}}$ are the latent space representations of the ${\bf x}_i$. While the mapping $\mathbb{P}_{XL}({\bf X})$ is obviously not bijective for $N_{\rm \ell}<N_\mathrm{f}$, it is crucial that the low-dimensional representation of the data in the latent space, ${\bf L}$, is able to address the essential correlations of the features of the original data ${\bf X}$. In this sense, it is desirable that the pairwise distance, $D (., .)$, in the feature space representation, ${\bf X}$, and in the latent space representation, ${\bf L}$, are conserved as good as possible, \ie, $D({\bf x}_i, {\bf x}_j) \simeq D ( {\bf l}_i, {\bf l}_j)$. The below described PCA is a very frequently used method for dimensional reduction which fulfills this requirement \cite{HIGHBIAS2019}. 

In describing the PCA we first assume -- without loss of generality -- zero empirical mean and unit variance of the ${\bf x}_i$ along the columns of ${\bf X}$. We consider the data set ${\bf X}$ as an ($N \times N_{\rm f}$)-matrix (also termed design matrix), whose rows are the $N$ data points and whose columns are the $N_{\rm f}$  features; we then construct the ($N_{\rm f} \times N_{\rm f}$), symmetric, positive-semidefinite covariance matrix, ${\mathbf \Sigma} ({\bf X})$, defined via

\begin{equation}
{\mathbf \Sigma} ({\bf X}) = \frac{1}{(N - 1)} {\bf X}^{\rm T} {\bf X} ;
\end{equation}
the superscript 'T' denotes the transpose of the matrix. The diagonal elements of ${\mathbf \Sigma} ({\bf X})$ measure the variance of features and the off-diagonal elements measure the covariance between features $i$ and $j$. ${\mathbf \Sigma} ({\bf X})$ can be diagonalized:

\begin{equation}
{\mathbf \Sigma} ({\bf X}) = {\bf V} {\mathbf \Lambda} {\bf V}^T ;
\end{equation}
it is assumed that the (real-valued, positive) eigenvalues $\lambda_i$, $i = 1, \dots, N_{\rm f}$ of the ($N_{\rm f} \times N_{\rm f}$) diagonal matrix ${\mathbf \Lambda}$ are arranged in descending order with respect to their value. The related eigenvectors are denoted by ${\bf v}_i$ and are collected in a matrix ${\bf V} = \{ {\bf v}_1, ..., {\bf v}_{N_{\rm f}} \}$. 

${\bf \Lambda}$ can now be used for dimensional reduction: large values of $\lambda_i$ label along the associated eigenvectors ${\bf v}_i$ directions of high variance in the feature space, which are, in many cases, those directions which contain the relevant information of the data. In contrast, directions associated with small values of $\lambda_i$ (\ie, with small variance) are usually related to noise and can potentially be ignored. The term ``principal component analysis'' originates exactly from the ordering $\lambda_1 > \lambda_2 > ... > \lambda_{N_{\rm f}}$ of the eigenvalues of ${\bf \Lambda}$: the eigenvector ${\bf v}_i$ with the largest (second largest, ...) eigenvalue $\lambda_i$ is referred to as the first (second, ...) principal component \cite{HIGHBIAS2019}, denoted in the following as PC.

Often, only very few of the $\lambda_i$ have a significant value. Selecting the $N_\ell$
largest eigenvalues and the associated eigenvectors, $\mathbf{V}^*=(\mathbf{v}_1, \dots , \mathbf{v}_{N_{\ell}}) \in \mathbb{R}^{N_{\rm f}\times N_{\ell}}$, provides us with an effective way to project the original data points into a low-dimensional (but representative) latent space $\mathbf{L}=\mathbb{P}_{XL}(\mathbf{X})$. In that way, the transformation $\mathbb{P}_{XL}$ is simply a linear projection from $\mathbb{R}^{N_f}$ onto $\mathbb{R}^{N_{\ell}}$ for every data point. For a more comprehensive discussion of the PCA see Ref.~\cite{Shlens2014}.

\subsubsection{$k$-means clustering}
\label{subsubsec:k_means}

Probably the simplest form of unsupervised learning is clustering algorithms, whose objective is to identify groups in unlabeled data according to similarity or distance measures of one kind or another~\cite{Moghadam2019, Boattini2019, HIGHBIAS2019}. In the following, we introduce the $k$-means algorithm~\cite{Steinhaus1956, Forgy1965, MacQueen1967, Lloyd1982} which we have used for our problem.

Starting again from $N$ data points, $\mathbf{X} = \{\mathbf{x}_1,\dotsc,\mathbf{x}_N\}$, in an $N_{\rm f}$-dimensional feature space, $\mathbf{x}_i\in\mathbb{R}^{N_{\rm f}}$, the objective is to distribute a certain number of $K$ cluster centers, called the cluster means $\mathbf{K}=\{\boldsymbol{\mu}_1,\boldsymbol{\mu}_2,\dotsc,\boldsymbol{\mu}_K\}$ with $\boldsymbol{\mu}_k\in\mathbb{R}^{N_{\rm f}}$, in the feature space, such that data points assigned to the different clusters minimize the following cost function

\begin{equation}
    \mathcal{C}(\mathbf{X}, \mathbf{K})=\sum_{k=1}^{K}\sum_{i=1}^{N}r_{ik}(\mathbf{x}_i-\boldsymbol{\mu}_k)^2.
    \label{eq:methods:order:clustering:kmeans}
\end{equation}
In this relation, the assignment of data point $i$ to cluster $k$ is realized via the binary variable $r_{ik}=1$ (and $r_{ik^\prime}=0$ for all $k^\prime\neq k$).
$\sum_{i=1}^{N}r_{ik}=N_k$ defines the size of cluster $k$, \ie, the number of data points associated with it. The set of assignments $\mathbf{k}=\{r_{ik}\}$ is also called labeling or clustering of the data points.

Minimizing Eq.~(\ref{eq:methods:order:clustering:kmeans}) can be interpreted as finding $\mathbf{K}$ and assigning  via the $r_{ik}$ the $N$ data points to different clusters, $k$, such that the (scaled) variance of each cluster, $\sum_{i=1}^{N}r_{ik}(\mathbf{x}_i-\boldsymbol{\mu}_k)^2$, is minimized. In practice this task is performed in a two-step procedure \cite{HIGHBIAS2019}:

\begin{itemize}
    \item[1.] Eq.~(\ref{eq:methods:order:clustering:kmeans}) is minimized with respect to $\boldsymbol{\mu}_k$ given a set of assignments $\{r_{ik}\}$, \ie, $(\partial\mathcal{C}/\partial\boldsymbol{\mu}_k)|_{\{r_{ik}\}}=0$, yielding the update rule for $\boldsymbol{\mu}_k = N_k^{-1}\sum_{i=1}^N r_{ik}\mathbf{x}_i$; thus $\boldsymbol{\mu}_k$ is the geometric center of the members $r_{ik}\mathbf{x}_i$ of cluster $k$;
    \item [2.] given the cluster means $\mathbf{K}$ we want to find the assignments $\mathbf{k}=\{r_{ik}\}$ which minimize Eq.~(\ref{eq:methods:order:clustering:kmeans}) by assigning each data point to its nearest cluster-mean: $r_{ik} = 1$ if $k=\arg [\min_{k^\prime}(\mathbf{x}_i - \boldsymbol{\mu}_{k^\prime})^2]$ and $r_{ik} = 0$ otherwise.
\end{itemize}

These two steps are performed in an alternating manner until some convergence criterion is met: this can, for instance, be the case if the change of the object function, given by Eq.~(\ref{eq:methods:order:clustering:kmeans}), between two iteration steps, is smaller than a predefined threshold value.

The $k$-means algorithm scales linearly with the size of the data set and can therefore be used for a large amount of data. However, Eq.~(\ref{eq:methods:order:clustering:kmeans}) is in general a non-convex function and the result for the minimization may largely depend on the initial (random) choice of the means $\mathbf{K}$ and the assignments $\mathbf{k}=\{r_{ik}\}$. In practice, the $k$-means algorithm is therefore applied several times with different (random) initial conditions which may result in different assignments (see discussion in Sec. \ref{subsec:asymmetric_bilayer} and Appendix \ref{app:reliability}). Eventually, the particular assignment with the minimal value of $\mathcal{C}(\mathbf{X},\mathbf{K})$ -- as compared to all other assignments -- is chosen to be the ``best'' solution to the clustering problem.



\section{Results}
\label{sec:results}

\subsection{The original diagram of states}
\label{subsec:original_diagram_of_states}

\begin{figure}[htb]
\begin{center}
\includegraphics[width=\textwidth,clip=True]{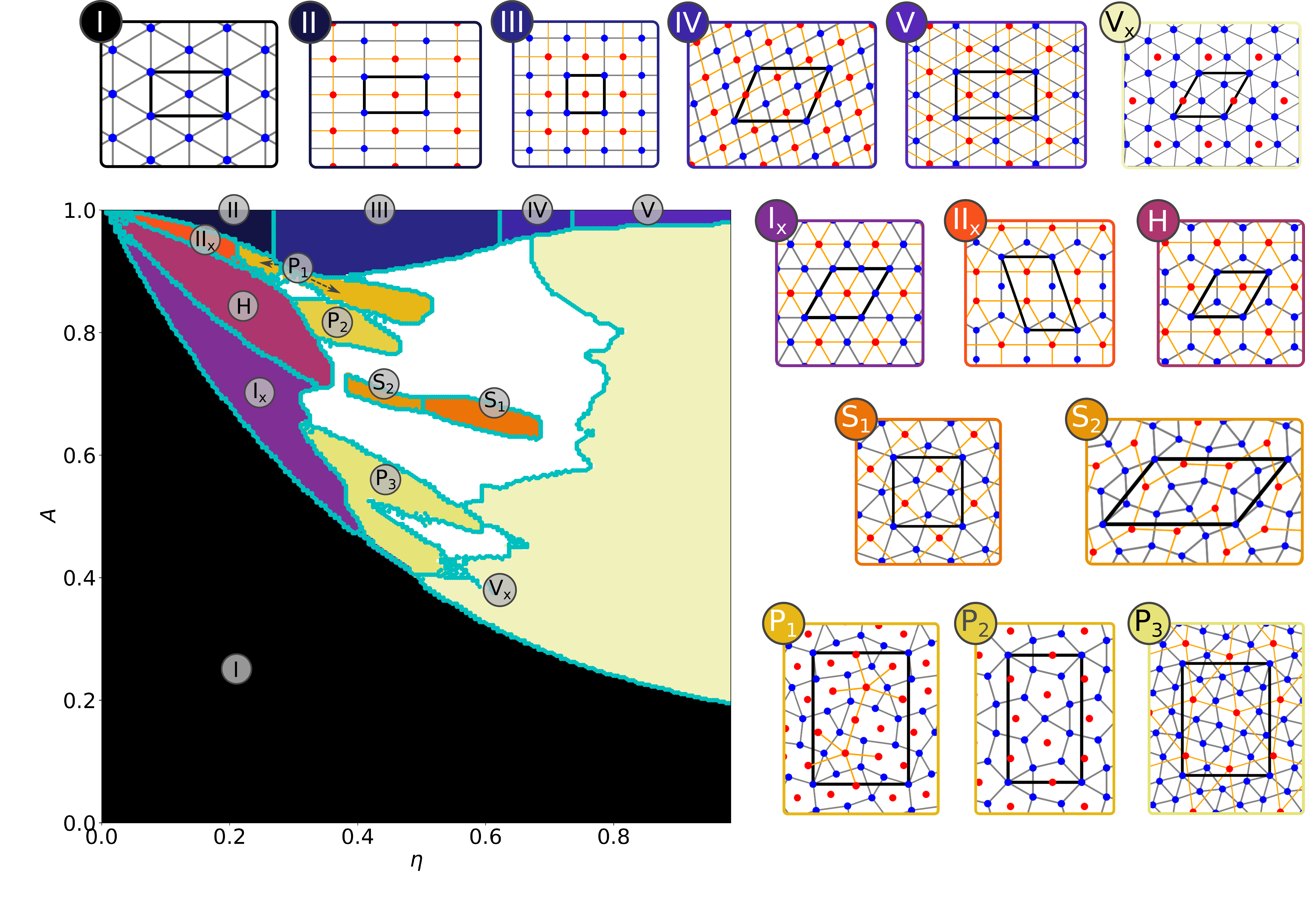}
\caption{Diagram of states of the ground state configurations of the asymmetric Wigner bilayer system in the $(A, \eta)$-plane, redrawn from Refs.~\cite{MoritzAntlanger2015, Antlanger2016, Antlanger2018}. Colored and labeled regions denote the fourteen phases that have been identified ``by hand'' according to the specifications summarized in Table \ref{tbl:labels}. The corresponding structures are shown in separate panels, specified by the same labeling and the same color code. In these panels, layer one (${\cal L}_1$) particles are colored in blue and layer two (${\cal L}_2$) particles are colored red; the thick black frames indicate the unit cells of the bilayer structures. The cyan lines in the main panel highlight the phase boundaries. Structures within the white region have not been classified, yet.}
\label{fig:phasediagram_antlanger}
\end{center}
\end{figure}

The original diagram of states of the asymmetric Wigner bilayer system, presented in Refs. \cite{MoritzAntlanger2015,Antlanger2016,Antlanger2018}, has been redrawn in Fig.~\ref{fig:phasediagram_antlanger}. The indicated fourteen ordered configurations have been identified ``by hand'' according to the criteria summarized in Table \ref{tbl:labels}, based on the composition $x$ and on the BOOPs $\Psi_{[4, 5, 6]}^{(1, 2, 3, 4)}$. From Fig. \ref{fig:phasediagram_antlanger} it is obvious that quite extended regions in the $(A, \eta)$-plane remain unidentified with this classification scheme.

\begin{table}[h!]
  \centering
  \begin{tabular}{l|l|l}
   ~ & characteristic features & composition and order parameters \\
   \hline
   \hline
   \I & hexagonal monolayer & $x=0$ \\
   \II & rectangular bilayer & $x=\frac{1}{2}$, $\Psi^{(1,2)}_4=1$, $0<\Psi^{(1,2)}_6<1$ \\
   \III & square bilayer & $x=\frac{1}{2}$, $\Psi^{(1,2)}_4=1$, $\Psi^{(1,2)}_6=0$ \\
   \IV & rhombic bilayer & $x=\frac{1}{2}$, $0<\Psi^{(1,2)}_4<1$, $0<\Psi^{(1,2)}_6<1$ \\
   \Vs & hexagonal bilayer & $x=\frac{1}{2}$, $\Psi^{(1,2)}_4=0$, $\Psi^{(1,2)}_6=1$ \\
   \hline
   \hline
   \Ix & trihexagonal (layer one) & $0 < x < \frac{1}{3}$, $\Psi_6^{(3)} > 0.9$ \\
   \Honey & honeycomb (layer one) & $x=\frac{1}{3}$, $\Psi_6^{(3)} > 0.9$ \\
   \IIx & modified rectangular bilayer & $\frac{1}{3}<x<\frac{1}{2}$, $\Psi_6^{(3)} > 0.9$ \\
   \Vx & hexagonal bilayer & $0 < x < \frac{A}{1+A}$, $(1-x)\Psi_6^{(1)} + x \Psi_6^{(2)} > 0.9$ \\
   \SnubSquare & snub square (layer one) & $x=\frac{2}{6}$, $\Psi_5^{(1)} > 0.7$, $\Psi_4^{(2)} > 0.9$ \\
   \hline
   \hline
   \SnubSquareTwo & snub square like (layer two) & $x=\frac{2}{6}$, $\Psi_5^{(2)} > 0.45$ \\
   \Pone & pentagonal type two & $\frac{1}{3}<x<\frac{1}{2}$, $\Psi_5^{(2)} > 0.45$ \\
   \Ptwo & pentagonal holes & $\frac{1}{3}<x<\frac{1}{2}$, $\Psi_5^{(4)} > 0.9$ \\
   \Pthree &  pentagonal holes & $0<x<\frac{1}{3}$, $\Psi_5^{(4)} > 0.9$ \\
   \hline
  \end{tabular}
  \caption{List of the fourteen ordered ground state configurations as identified in Refs. \cite{MoritzAntlanger2015, Antlanger2016, Antlanger2018} ``by hand'' (see also Fig.~\ref{fig:phasediagram_antlanger}): shorthand notation of the respective configurations (left column), short description of the characteristic features (central column), and specification in terms of the composition of the system, $x$, and order parameters $\Psi_\nu^{(u)}$ (right column). Note that all structures -- except for structure \I~ which refers to a monolayer -- are bilayer structures.}
  \label{tbl:labels}
\end{table}

\subsection{Unsupervised clustering algorithms applied to the Wigner bilayer}
\label{subsec:unsupervised_wigner}

The huge amount of data accumulated  in Refs. \cite{MoritzAntlanger2015, Antlanger2016, Antlanger2018} in the search for equilibrium structures and the fact that -- despite considerable efforts -- white regions still remain in the diagram of states has motivated us to re-visit and to re-analyze this data set  with the help of a tool that is able to perform such a classification in a more systematic and more efficient manner as one could possibly do ``by hand''. To this end, we have introduced a more systematic classification scheme based on unsupervised clustering (as introduced in Sec. \ref{subsec:unsupervised_clustering}).

In order to apply this unsupervised clustering algorithm scheme we first have to define the $N_{\rm f}$-dimensional feature vector ${\bf x}(\Genome)$, which is built up by the following components:

\begin{itemize}
    \item the set of BOOPs, originally used in \cite{MoritzAntlanger2015, Antlanger2016, Antlanger2018}, $\Psi_{[4, 5, 6]}^{(1, 2, 3, 4)}$, has been extended by the set $\Psi_{[3, 8, 10, 12]}^{(1, 2, 3, 4)}$ (using the short-hand notation introduced above); this new set of in total 28 BOOPs offers now the possibility to identify structures with eightfold, tenfold, or twelvefold symmetries; 
    \item the composition $x = x(\Genome)$; 
    \item in an effort to quantify the ratio of the average nearest neighbor distance in ${\cal L}_1$, $r_\mathrm{nn}^{(1)}(\Genome)$, and the average nearest neighbor distances in ${\cal L}_2$, $r_\mathrm{nn}^{(2)}(\Genome)$, for a certain bilayer configuration, $\Genome$, we define the ``intralayer nearest neighbor ratio'' order parameter, $r_g(\Genome)$, via
    
    \begin{equation}
    r_g(\Genome)=\frac{r_\mathrm{nn}^{(1)}(\Genome)}{r_\mathrm{nn}^{(2)}(\Genome)} .
    \label{eq:arg_radial}
\end{equation}
The values of $r_g(\Genome)$ are not bound to a maximum value; however, we find empirically an upper limit of $\simeq 1.07$ for all considered bilayer ground state configurations.
\end{itemize}
    
With the 28 BOOPs, the values for $x$ and $r_g(\Genome)$ we end up with a feature vector that has $N_{\rm f} = 30$ components $f_i(\Genome)$: 

\begin{equation}
  \mathbf{x}(\Genome) = \{f_1(\Genome),\dotsc,f_{N_{\rm f}=30}(\Genome)\} .
  \label{eq:wigner:features}
\end{equation}

In the following, we will re-analyze the classification scheme of phases used in  Refs.~\cite{MoritzAntlanger2015, Antlanger2016, Antlanger2018} with the help of unsupervised machine learning techniques in order to automatically identify different families of structures directly from the feature vector, $\mathbf{x}$, given in Eq.~(\ref{eq:wigner:features}). To be more specific: {\bf (i)} we first perform a principal component analysis \cite{Jolliffe2002} (PCA) on the feature vectors of all structures from Ref.~\cite{MoritzAntlanger2015, Antlanger2016, Antlanger2018}, which defines our basic data set. This allows us to identify directions of large variance in the data set which capture the most relevant information among the different features. To this end we transform the data set of feature vectors to unit-variance and zero-mean coordinates; this technique is termed ``whitening'' in literature and is used to decouple the PCA from the relative scales of different features~\cite{HIGHBIAS2019}.  {\bf (ii)} We then apply the $k$-means \cite{Steinhaus1956, Forgy1965, MacQueen1967, Lloyd1982} clustering algorithm to the latent space representation of the data set which is spanned by the leading PCs. This will help us to identify new, previously unclassified phases which are potentially hidden in the huge set of the original structural data.

As a benchmark for this approach and for illustrative reasons we start our analysis with the simplest problem within the topic of Wigner bilayers, namely with the {\it symmetric} Wigner bilayer system (as considered in Refs.~\cite{Samaj2012, Samaj2012a}) where $\sigma_1=\sigma_2$ or, equivalently, $A \equiv 1$.

\subsection{The symmetric Wigner bilayer system -- a testing case}
\label{subsec:revisited_diagram_of_states}

\begin{figure}[htb]
\begin{center}
  \includegraphics[width=\textwidth, clip=True]{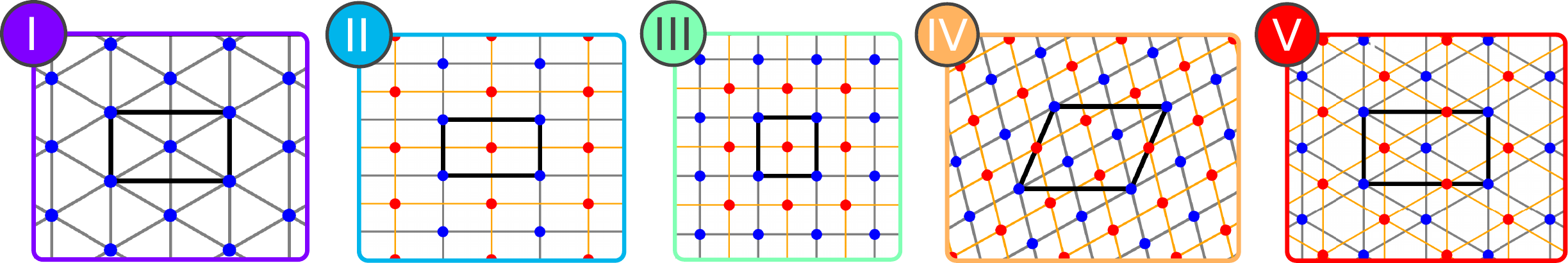}
  \caption{Phases {\I} through {\Vs} (as color-coded) representing the ground state configurations of the symmetric Wigner bilayer system \cite{Samaj2012, Samaj2012a}. Blue and red symbols represent particles from layers ${\cal L}_1$ and ${\cal L}_2$, respectively. The respective unit cells are indicated by black frames.}
\label{fig:wigner:symmetric:states}
\end{center}
\end{figure}

For the symmetric case, the identification of the ground state configurations has been solved analytically \cite{Samaj2012, Samaj2012a}, with five emerging structures, labeled {\I} through {\Vs}; these phases are  depicted in Fig.~\ref{fig:wigner:symmetric:states}. Furthermore, the exact $\eta$-values where the transitions between these phases occur as well as the nature of these transitions could be identified with high accuracy in the above contributions: the hexagonal monolayer (\I) is stable only at $\eta=0$ and transforms for an infinitesimally small value of $\eta$ into a rectangular bilayer, termed \II. 
This structure is stable within the range $0 < \eta \lesssim 0.263$ and then transforms via a second-order transition into a square bilayer (\III), which is stable within the range $0.263 \lesssim \eta \lesssim0.621$. This structure then turns -- again via a second order transition -- into a rhombic bilayer phase (\IV), stable within  $0.621 < \eta \leq 0.728$. Eventually, a hexagonal bilayer (\Vs) emerges at $\eta \simeq 0.728$ via a first-order transition (see also the line $(A=1)$ in Fig.~\ref{fig:phasediagram_antlanger}).

We now test the clustering approach for this particular case where we have the solution already at hand. These calculations are based on the ${N_\mathrm{sym}}=141$ ground state configurations that were identified via the memetic evolutionary algorithm  in Refs.~\cite{MoritzAntlanger2015, Antlanger2016, Antlanger2018} for different values of $\eta \in [0,1]$ and for $A \equiv 1$.

We first perform for all structures a principal component analysis (PCA) \cite{Jolliffe2002} on the set of the (unit-variance and zero-mean) feature vectors, $\mathbf{X}^\mathrm{(sym)}$, defined in Eq.~(\ref{eq:wigner:features}) with $N_{\rm f} = 30$. The data set is then transformed into an $N_{\rm \ell}$-dimensional latent space representation, $\mathbf{L}^\mathrm{(sym)}=\{\mathbf{l}_1,\dotsc,\mathbf{l}_{N_\mathrm{sym}}\}$ with $\mathbf{l}_i=(v_{i1},v_{i2},\dotsc,v_{i N_{\rm \ell}})\in\mathbb{R}^{N_{\rm \ell}}$ and $N_{\rm \ell} \leq N_{\rm f}$. The actual value of $N_{\rm \ell}$ defines how many leading PCs are considered in the latent space representation of the data.

When investigating the $N_{\rm sym}$ data points as a function of the first three PCs (corresponding to the data points ${\bf x}_i$ projected onto the first three latent space directions  $\mathbf{v}_1$, $\mathbf{v}_2$, and $\mathbf{v}_3$) one can already distinguish the different phases by eye: structures belonging to a specific phase form clusters in such a representation which are spatially separated from each other (data not shown here, cf. top panel in Fig.~3.5 of Ref. \cite{Hartl2020Thesis}).  

\begin{figure}[htb]
\begin{center}
  \includegraphics[width=\textwidth, clip=True]{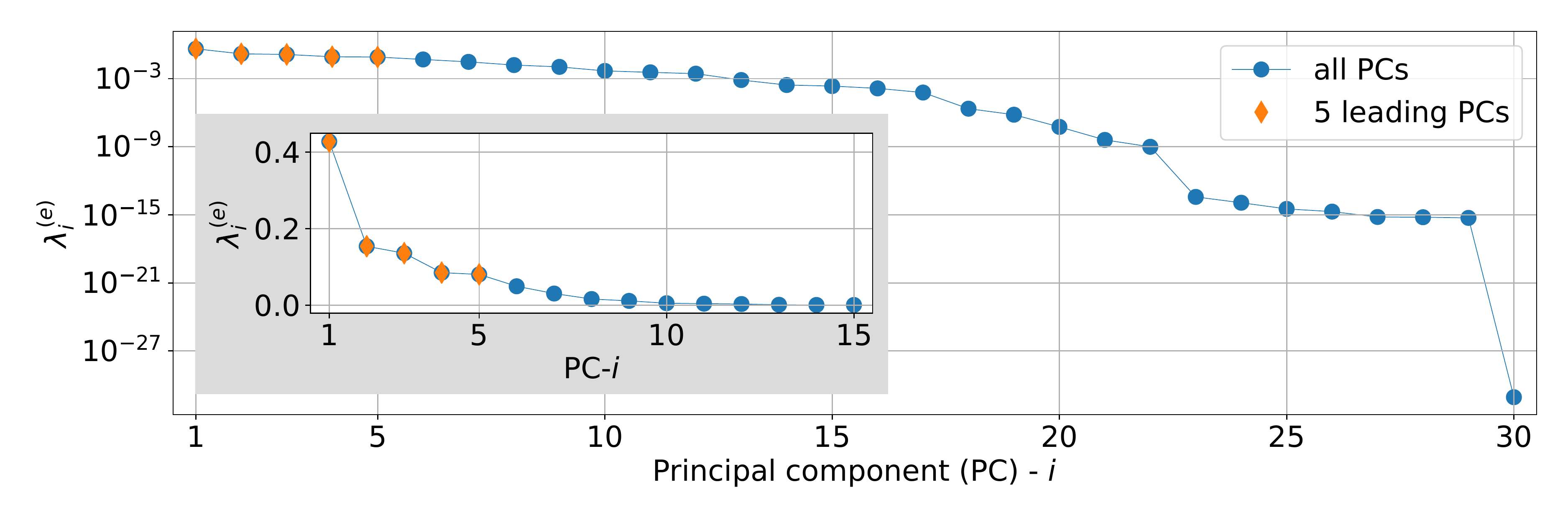}
  \caption{Percentage of the explained variance, $\lambda_i^{\rm (e)}$, for each PC $i$ as defined in Eq.~(\ref{eq:explained_variance}) for all 30 PCs (blue) of the symmetric Wigner bilayer system. The leading five principal components are shown in orange. Inset: same data as in the main plot but with linear scale for $\lambda_i^{\rm (e)}$ for the first 15 PCs.  }
\label{fig:wigner:symmetric:pca}
\end{center}
\end{figure}

In Fig.~\ref{fig:wigner:symmetric:pca} we present the percentage of the explained variance (PEV), $\lambda_i^{\mathrm{(e)}}$, contained in each PC with index $i$, defined as

\begin{equation}
    \lambda_i^{\rm (e)} = \left( \sum_{j=1}^{N_{\rm f}} \lambda_j \right)^{-1} \lambda_i.
    \label{eq:explained_variance}
    \end{equation}
The PEV quantifies the amount of information encoded in each PC direction $\mathbf{v}_i$. We see that the values of $\lambda_i^{\mathrm{(e)}}$ quickly drop from $\lambda_1^{\mathrm{(e)}}\sim1/3$ to $\lambda_6^{\mathrm{(e)}}< 0.05$, and furtheron by several orders of magnitudes, such that the higher PC (\ie, for $i \gtrsim 6$) are insignificant as compared to the leading ones. Thus, we can safely restrict ourselves to the five leading PCs and set in the following $N_\ell=5$.

We now apply the $k$-means clustering algorithm (cf. Subsec. \ref{subsubsec:general_remarks}) to the ($N_\ell =5$)-dimensional latent space representation $\mathbf{L}^\mathrm{(sym)}$ of the data $\mathbf{X}^\mathrm{(sym)}$ and assign to all $i=1,\dotsc,{N_\mathrm{sym}}$ data points a cluster label $c_i\in\{1,\dotsc,K\}$, defining thereby the labeling (or clustering) $\mathbf{k}^\mathrm{(sym)}=\{c_1, \dotsc,c_{N_\mathrm{sym}}\}$ of the data set. In the particular case of the symmetric Wigner bilayer system we already know the numbers of phases and therefore set $K=5$. 

\begin{figure}[htb]
\begin{center}
  \includegraphics[width=0.8\textwidth]{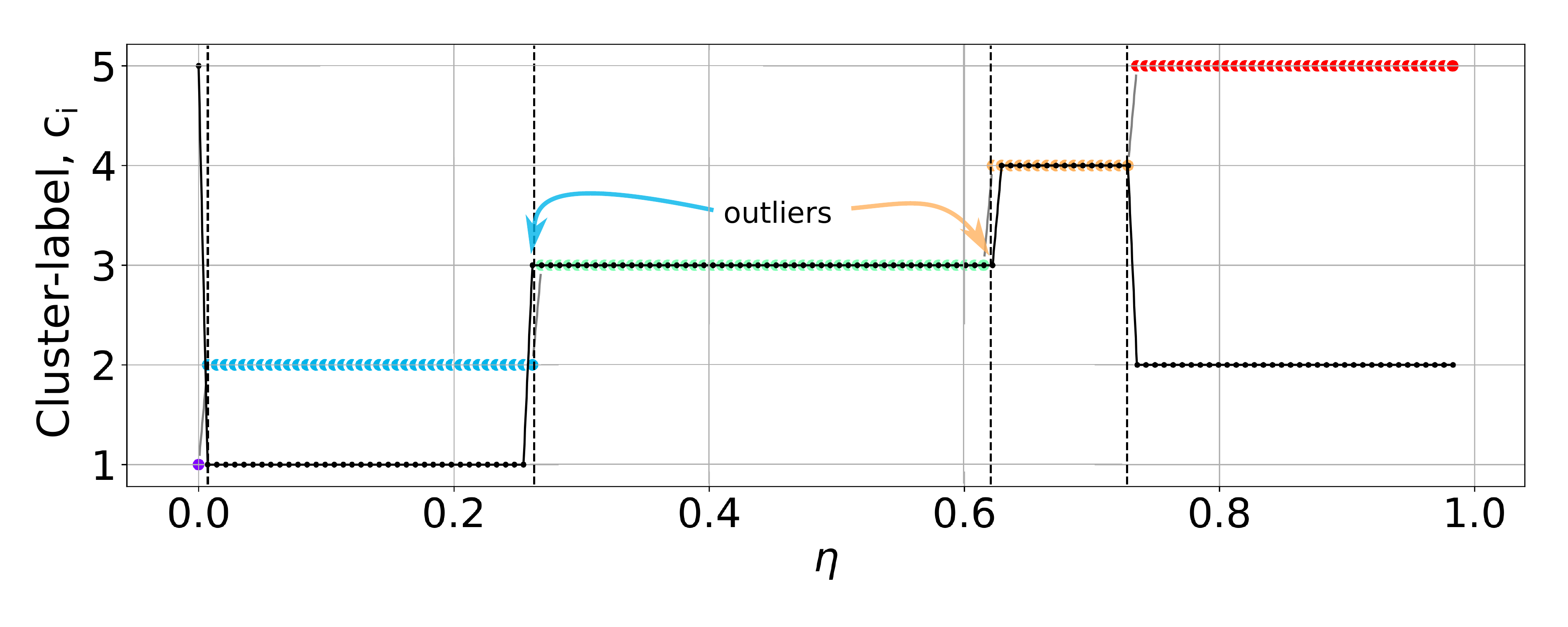}
  \caption{Labeling $\mathbf{w}^{(\mathrm{sym})}$~\cite{Samaj2012, Samaj2012a} of the ground state configurations of the symmetric Wigner bilayer system, \ie, \I: 1, \II: 2, \III: 3, \IV: 4 and \Vs: 5 (with symbols color-coded according to Fig.~\ref{fig:wigner:symmetric:states}), and labeling $\mathbf{k}^\mathrm{(sym)}$ by the $k$-means clustering (black symbols) for each of the ${N_\mathrm{sym}}$ data points of the data set $\mathbf{X}^\mathrm{(sym)}$ identified in Ref.~\cite{MoritzAntlanger2015} for different values of $\eta\in[0,1]$ at $A=1$. Note that the numerical value of a particular cluster label $c_i$ (= 1 through 5) assigned by the $k$-means algorithm to all data points belonging to one particular cluster is arbitrary but unique. An adjusted mutual information score, 
  $I_K(\mathbf{k}^\mathrm{(sym)},\mathbf{w}^{(\mathrm{sym})})=0.95$
  (as defined in Eq.~(\ref{eq:methods:order:mutual_information}) of Appendix \ref{app:adjusted_mutual_information}) is realized. The two outliers (highlighted by the arrows) are discussed in the text. 
  The vertical dashed lines mark the $\eta$-values where phase transitions from phase \I~through \Vs~occur: from left to right, $\eta=1/141$ represents the emergence of phase \II~(given the discrete steps in $\eta$), while  $\eta=0.263$, $\eta=0.621$, and $\eta=0.728$ mark the transitions from phase \II~to \III, \III~to \IV, and  \IV~to \Vs, respectively.}
\label{fig:wigner:symmetric:clustering}
\end{center}
\end{figure}

Results are shown in Fig.~\ref{fig:wigner:symmetric:clustering}. It can be seen that the emerging $k$-means clustering $\mathbf{k}^\mathrm{(sym)}$ of the data is in excellent agreement with the phase-assignment known from literature \cite{Samaj2012, Samaj2012a}, $\mathbf{w}^{(\mathrm{sym})}=\{C_1,\dotsc,C_{N_\mathrm{sym}}\}$; here the $C_i$ (=1 through 5) label the corresponding phases (\I~ through \Vs), respectively, for every data point $i$ (as specified in Table \ref{tbl:labels}). It should be noted that the particular numerical values that associate the data points with a certain cluster are usually arbitrarily chosen by the $k$-means algorithm; however they are unique: in Fig.~\ref{fig:wigner:symmetric:clustering} we see that the clusters related to phases \I~ through \Vs~ are, respectively, labeled by $c_i=5,1,3,4$ and $2$. Further, the assignment of the data points into the different clusters is almost perfect and the labels $c_i$ can be redefined to match the numerical values of $C_i$ (not shown here). 

In an effort to test the reliability of the partitioning of a data set, we use the so-called mutual information score, introduced and discussed in Appendix \ref{app:adjusted_mutual_information}. For this particular case, we report an (adjusted) mutual information score of 
$I_K(\mathbf{k}^\mathrm{(sym)}, \mathbf{w}^{(\mathrm{sym})})=0.95$, 
as defined in Eq.~(\ref{eq:methods:order:mutual_information}) of Appendix \ref{app:adjusted_mutual_information},  between the clustering $\mathbf{k}^\mathrm{(sym)}$ and the phase-assignment from literature $\mathbf{w}^{(\mathrm{sym})}$~\cite{Samaj2012, Samaj2012a}. The small discrepancies arising from two data points (denoted in Fig.~\ref{fig:wigner:symmetric:clustering} as \textit{outliers}) are discussed below.

Usually, the results from the $k$-means clustering depend on the initial conditions of the algorithm such as (i) the initial (and usually arbitrary) placement of the $K$ different cluster centers in the latent space of the data set $\mathbf{X}^\mathrm{(sym)}$ and (ii) the initial data point assignments to the clusters. To justify the results shown in Fig.~\ref{fig:wigner:symmetric:clustering} we thus perform 100 independent runs of $k$-means clustering (labeled with an index $l$) on the leading five principal components of the data set $\mathbf{X}^\mathrm{(sym)}$. We find that all corresponding labelings, $\mathbf{k}^\mathrm{(sym)}_l$, share the same adjusted mutual information score,  
$I_K(\mathbf{k}^\mathrm{(sym)}_l,\mathbf{w}^{(\mathrm{sym})})=0.95$ 
with the results $\mathbf{w}^{(\mathrm{sym})}$ being known from literature . 

Furthermore, we also increased the number of leading PCs from five to 30 without observing significant changes in the results; however, when using less than five principal components the results become unreliable. These observations confirm that the clustering shown in Fig.~\ref{fig:wigner:symmetric:clustering} represents now indeed the optimal $k$-means clustering to group the data points of $\mathbf{X}^\mathrm{(sym)}$ into the phases \I~ through \Vs.

In Fig.~\ref{fig:wigner:symmetric:clustering} two outlier structures are highlighted by arrows which are located at $\eta$-values close to transition boundaries from phase \II~ to phase \III~ as well as from phase \III~ to phase \IV, respectively. The reason that these structures are -- erroneously -- attributed by the $k$-means algorithm to phase \III~ lies in the fact that for the respective $\eta$-values both a rectangular bilayer structure (phase \II) or a rhombic bilayer structure (phase \IV) can righteously be considered (within numerical accuracy) as ``nearly''-square structures (phase \III).

We point out that small numerical variations in the data, which are often related to artifacts (such as noise), can trigger undesired effects in clustering approaches and may lead to an artificial partitioning of data in a clustering or classification task. Therefore a proper preparation of the data with, for instance, PCA can help to reduce the effects of noise on the outcome of a clustering approach of a particular data set. On the other hand, sometimes small variations in the data do have a physical meaning such as, for instance when continuous phase transitions occur; in such a case particular caution has to be taken for correctly distinguishing between different clusters of data points.

\subsection{The asymmetric Wigner bilayer system}
\label{subsec:asymmetric_bilayer}

Now that we know from the {\it symmetric} case of the Wigner bilayer system we can firmly rely on the analysis approaches detailed in Subsec. \ref{subsec:unsupervised_clustering}, we proceed to the {\it asymmetric} case;  this structural database was generated for the preceding contributions \cite{MoritzAntlanger2015, Antlanger2016, Antlanger2018} by independent evolutionary structure optimization of configurations with up to 40 particles per unit cell and considering all related possible values of the composition, $x$, on an $A$- and $\eta$-grid as specified in Subsec. \ref{subsec:lattice_sum}. We  perform the same analysis -- \ie, first a PCA and then a subsequent $k$-means clustering -- on the set of feature vectors, $\mathbf{X}^\mathrm{(asym)}=(\mathbf{x}_1,\cdots,\mathbf{x}_{N_\mathrm{asym}})$ with the  $\mathbf{x}_i\in\mathbb{R}^{N_f=30}$ being taken initially from the entire set of $N_\mathrm{asym} \sim 56541$ configurations considered in the asymmetric Wigner bilayer system. 

\paragraph{Principal component analysis (PCA)}

\begin{figure}[htb]
\begin{center}
    \includegraphics[width=\textwidth, clip=True]{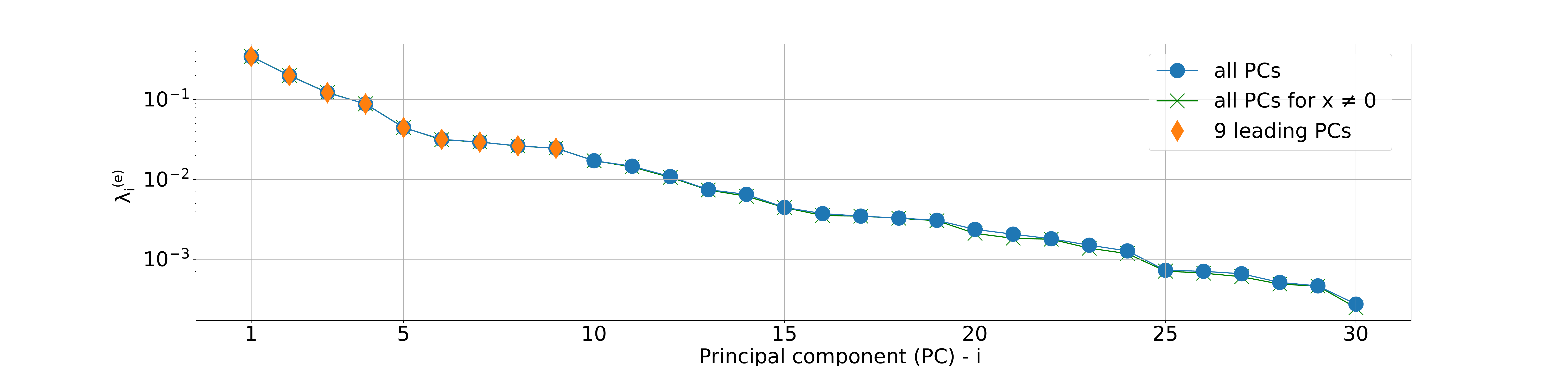}
    \caption{Percentage of the explained variance, (PEV), $\lambda_i^{\mathrm{(e)}}$ (as defined in Eq.~(\ref{eq:explained_variance})), of the principal components (PCs) of the data, set $\mathbf{X}^\mathrm{(asym)}$ of feature vectors of all $N_\mathrm{asym}\sim 56541$ configurations of the asymmetric Wigner bilayer system, considered in Refs.~\cite{MoritzAntlanger2015, Antlanger2016, Antlanger2018} (blue dots); the leading nine PCs (with the PEV $\lambda_j^{\mathrm{(e)}} > 0.02$) are shown in orange. We also present (via green crosses) the PEV of the PCs of the data set $\mathbf{X}^\mathrm{(*)}$, \ie, the data set which contains all data points of $\mathbf{X}^\mathrm{(asym)}$ except for those which correspond to hexagonal monolayer configurations (\ie, where $x=N_2/N=0$).
    It should be noted that structures with $x=0$ can uniquely be characterized as trigonal monolayers, \ie, phase \I, in the investigated data set.
    Thus, the PEV results emphasize that the PCA is only marginally affected by the large proportion of phase \I~ structures in the data set.
    }
    \label{fig:wigner:phasediagram:clustering:pev}
\end{center}
\end{figure}

Via the PCA we first transform the data set $\mathbf{X}^\mathrm{(asym)}$ to a latent space representation $\mathbf{L}^\mathrm{(asym)}=(\mathbf{l}_1,\dotsc,\mathbf{l}_{N_\mathrm{asym}})$ of the data (for which we again assume unit-variance and zero-mean coordinates):  $\mathbf{l}_i(\in\mathbb{R}^{N_\ell})$ is the projection of the data point $\mathbf{x}_i\in\mathbb{R}^{N_f}$ into the latent space of dimension $N_\ell(\leq N_f)$;  note that the actual value of $N_\ell$ has not been fixed, yet.  In Fig.~\ref{fig:wigner:phasediagram:clustering:pev} we present the PEVs, $\lambda_j^{\mathrm{(e)}}$, defined in Eq.~(\ref{eq:explained_variance}), for each of the $N_{\rm f} = 30$ PCs of the feature vectors, $\mathbf{X}^\mathrm{(asym)}$, of the  structures considered in Refs.~\cite{MoritzAntlanger2015,Antlanger2016,Antlanger2018}. We conclude from the PEVs that -- similar to the symmetric case  (cf. Fig.~\ref{fig:wigner:symmetric:pca}) -- only very few principal components are expected to carry relevant information, \ie, will have significant $\lambda_i^{\mathrm{(e)}}$-values. Thus, we restrict our analysis to the leading nine principal components, whose $\lambda_i^{\mathrm{(e)}}$-values are larger than 0.02; hence we set $N_\ell=9$ in what follows.

\begin{figure}[ht]
\begin{center}
    \includegraphics[width=\textwidth, clip=True]{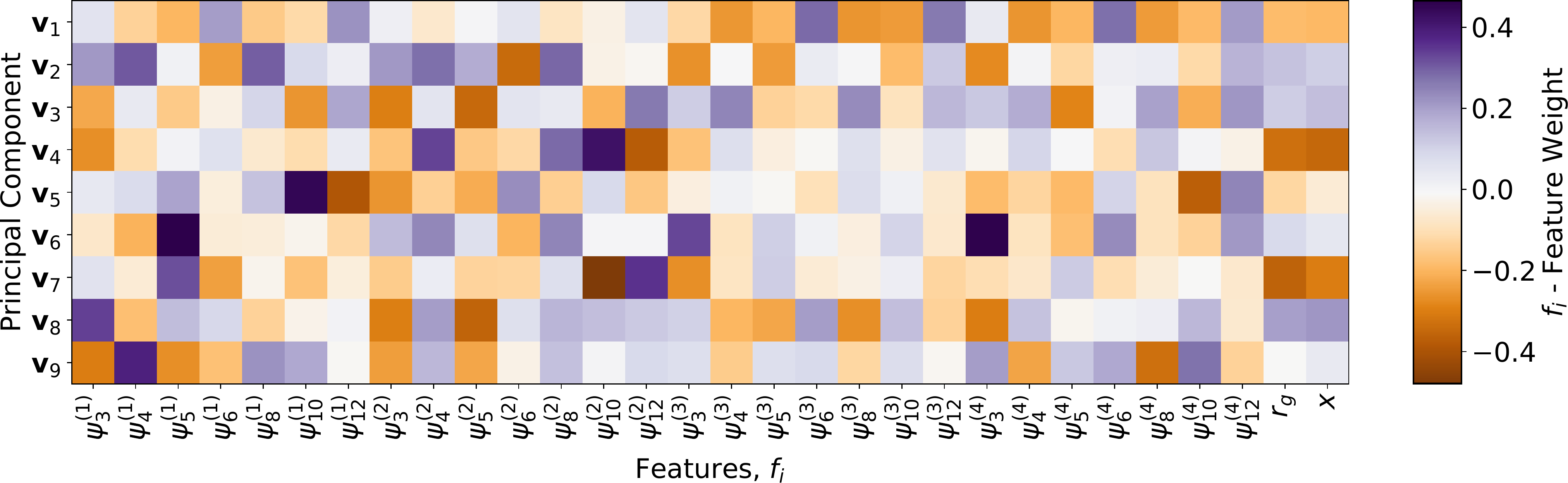}
    \caption{Leading nine PCs, $\mathbf{v}_1,\dotsc,\mathbf{v}_9\in\mathbb{R}^{N_f=30}$, of the (whitened) data set $\mathbf{X}^\mathrm{(asym)}$ represented in the feature space spanning $\mathbf{x}=(f_1,\dotsc,f_{N_f})$, cf. Eq.~(\ref{eq:wigner:features}).
    The features $f_1,\dotsc,f_{30}$ are listed along the horizontal axis and represent the order parameters used to evaluate $\mathbf{X}^\mathrm{(asym)}$ from the structural database of the asymmetric Wigner bilayer system from literature~\cite{MoritzAntlanger2015, Antlanger2016, Antlanger2018}; the respective $30$ elements of the PCs, $\mathbf{v}_j$, related to the directions of $f_j$ are presented via the color-coding specified by the color bar on the right-hand side: large positive (negative) values of feature weights are emphasized by dark purple or orange coloring, while values close to zero are colored in white.
    It should be noted that prior to the PCA, all features are separately scaled to zero mean and unit variance (i.e., the data is ``whitened''). Thus, even positive valued features such as $r_g$ and $x$ can have negative feature weights in the corresponding principal components.}
\label{fig:wigner:clustering:principal_components}
\end{center}
\end{figure}

In Fig.~\ref{fig:wigner:clustering:principal_components} we present the leading nine PCs, $\mathbf{v}_1,\dotsc,\mathbf{v}_9$, which are vectors in the feature space spanned by $\mathbf{x}=(f_1,\dotsc,f_{N_f})$. The 30 elements of each PC indicate the direction of the PC in feature space; we refer to the values of these elements of PCs as \textit{feature weights}. Large positive or large negative values of certain feature weights of a particular PC indicate important features which quantify information in the data set with directions of high variance. These characteristic features of PCs can be used to identify important order parameters - or combinations of order parameters if several feature weights are dominant in a particular PC. On the other hand, feature weights close to zero indicate less relevant directions.

From the data presented in Fig.~\ref{fig:wigner:clustering:principal_components} we see that the first PC, $\mathbf{v}_1$, exhibits large positive feature weights of the six and twelvefold order parameters $\Psi^{(1,3,4)}_{[6, 12]}$ and medium to large negative weights of $\Psi^{(1,3,4)}_{[4,5,8,10]}$, $r_g$, and $x$. The  PCs $\mathbf{v}_2$ and $\mathbf{v}_3$ exhibit medium to large feature weights distributed over a range of order parameters which makes them more difficult to interpret than the feature weights of $\mathbf{v}_1$. From the fourth PC ($\mathbf{v}_4$) onwards single (or very few) directions in feature-space are relevant:  this applies, for instance, for $\Psi^{(2)}_{[10,12]}$ in the case of $\mathbf{v}_4$ and for $\Psi^{(1)}_{[10,12]}$ in the case of $\mathbf{v}_5$. Further, we learn from Fig.~\ref{fig:wigner:clustering:principal_components} that for all PCs the feature weights for the intralayer nearest neighbor order parameter, $r_g$, and for the composition, $x$, are strongly correlated. Thus, one could interpret our results such that the nearest neighbor distances, $r_g$, of particles in both layers are largely governed by the composition $x$ for low energy configurations of the system. This, in turn, might suggest that the particles tend to be distributed in ground state configurations of the system as uniformly as possible (constraint by the $(A, \eta)$-specific lattice formation) in both layers.

\begin{figure}[htb]
\begin{center}
\includegraphics[width=0.5\columnwidth, clip=True]{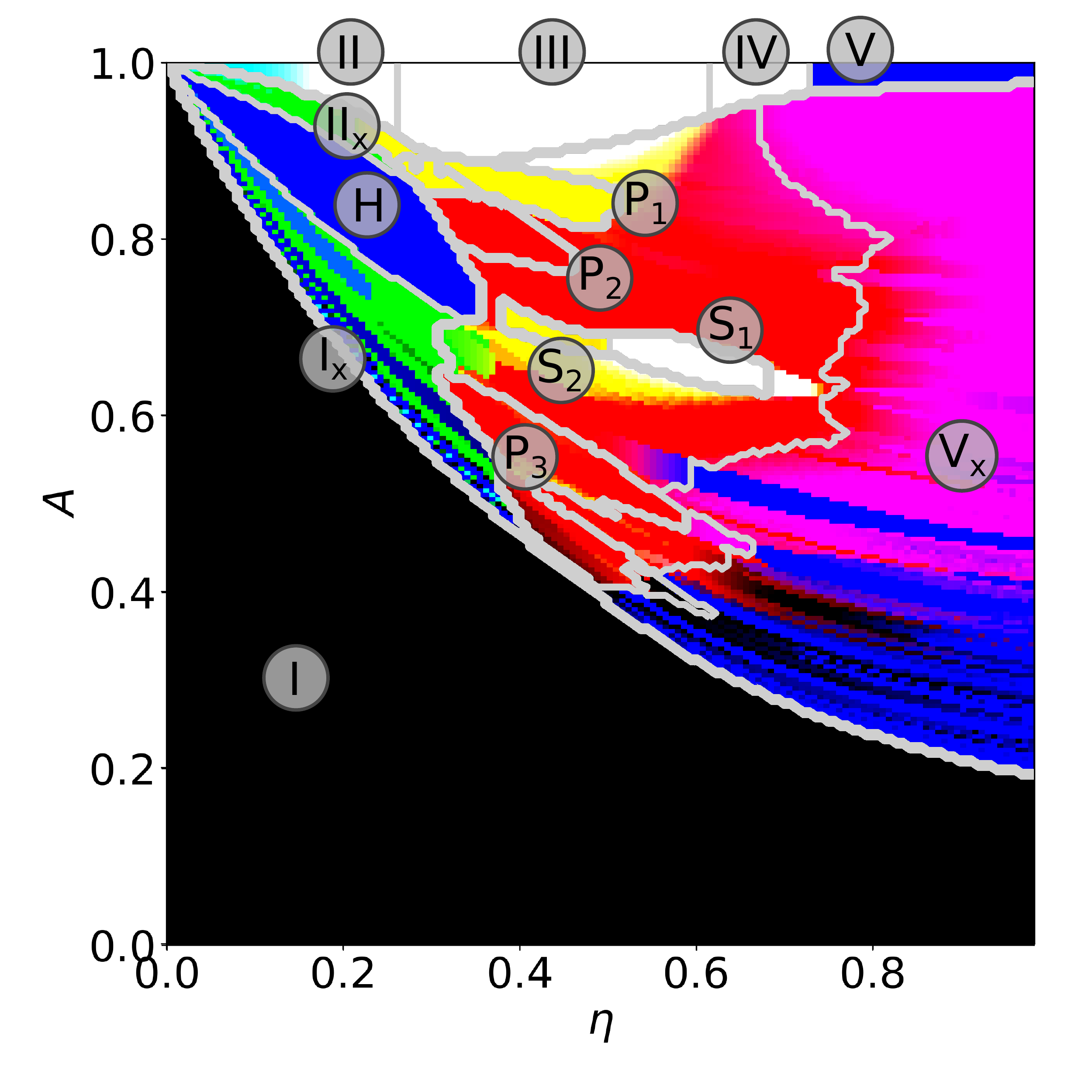}  
\caption{Diagram of states of the ground state configuration of the asymmetric Wigner bilayer system in the $(A, \eta)$-plane, coloured via an [R,~G,~B]-scheme that is based on the first three PC vectors $\mathbf{v}_1, \mathbf{v}_2$, and $\mathbf{v}_3$ of the data set $\mathbf{X}^\mathrm{(asym)}$: for every $(A, \eta)$-pair we define the relative amount of red, green, and blue color -- [R,~G,~B] -- of the corresponding pixel in the $(A, \eta)$-plane by the coordinates, $[v_{g1}, v_{g2}, v_{g3}]$, of the latent space data point, $\mathbf{l}_g=(v_{g1},\dotsc,v_{gN_\ell})$, of the associated ground state configuration of the asymmetric Wigner bilayer system, using the database available in literature \cite{MoritzAntlanger2015,Antlanger2016,Antlanger2018}; the values of the coordinates in the [R,~G,~B]-scheme are limited to the interval $[0, 1]$. The light gray lines indicate phase boundaries taken from Ref.~\cite{MoritzAntlanger2015, Antlanger2016, Antlanger2018}. Phases from the above references are labeled according to the classification of the 14 phases summarized in Table~\ref{tbl:labels}.}
\label{fig:wigner:phasediagram:clustering:pca}
\end{center}
\end{figure}

Based on this PCA we present in Fig.~\ref{fig:wigner:phasediagram:clustering:pca} the revised diagram of states of the ground state configurations of the asymmetric Wigner bilayer system in the $(A, \eta)$-plane in a new [R,~G,~B]-scheme, which is now based on the leading three PCs: to this end, we consider the latent space representations, $\mathbf{l}_g$, of the feature vectors, $\mathbf{x}_g$, which correspond to the suggested ground state configurations, $\Genome_g$, of the asymmetric Wigner bilayer system for different values of the system parameters, $A$ and $\eta$. For each of these data points, $\mathbf{l}_g=(v_{g1},\dotsc,v_{g N_\ell})$, we use the first three coordinates, $[v_{g1},v_{g2},v_{g3}]$, \ie, the coordinates of $\mathbf{l}_g$ associated to the first three PCs $\mathbf{v}_1$, $\mathbf{v}_2$ and $\mathbf{v}_3$, to define the relative contribution of the colors red, green, and blue to the color of each pixel in the $(A, \eta)$-plane. We can see that the phase boundaries as suggested in Refs.~\cite{MoritzAntlanger2015, Antlanger2016, Antlanger2018} nicely correlate with the values of the PCs; however, we can also spot out regions in the $(A, \eta)$-plane which call for a closer inspection: for instance, the region of phase \Ix~ is likely to have a more sophisticated internal structure than previously assumed, as indicated by the different greenish (\ie, dominant $\mathbf{v}_2$) and bluish (\ie, dominant $\mathbf{v}_3$) regions, a feature which we will further investigate in the following.

We can see from Fig.~\ref{fig:wigner:phasediagram:clustering:pca} that phase \I~ can uniquely be identified via the black color. Further, the leading three principal components of structures \II, \IIx, \Honey, and \Ix~ are clearly different from those of the pentagonal structures \Pone, \Ptwo, \Pthree,~ and \SnubSquareTwo:  (i) the former structures (\ie~  \II, \IIx, \Honey, and \Ix) are characterized by large [G, B]-values (associated to the second and third PCs), indicating thus large values of the latent space coordinates into directions $\mathbf{v}_2$ and $\mathbf{v}_3$; the corresponding bilayer structures have the property that -- when projecting the particle positions onto a single plate -- a hexagonal monolayer is formed; (ii) in contrast, the pentagonal structures (\Pone, \Ptwo, and \Pthree) are throughout more complicated to interpret. Still, we can see that their symmetry (cf. Fig.~\ref{fig:wigner:clustering:principal_components}) is either dominated by the first principal component $\mathbf{v}_1$ (red) or by combinations of $\mathbf{v}_1$ and $\mathbf{v}_2$ (yellow, which is generated by adding red and green in the [R,~G,~B]-notation). Furthermore, structures in the \Vx~ region show strong signals either from the third principal component $\mathbf{v_3}$ (blue) or from combinations of $\mathbf{v}_1$ and $\mathbf{v}_3$ (where red and blue become magenta); in that way they can be distinguished from the (red and yellow) pentagonal region in the diagram of states depicted in Fig.~\ref{fig:wigner:phasediagram:clustering:pca}. Summarizing we note that the region in the diagram of states depicted in Fig.~\ref{fig:wigner:phasediagram:clustering:pca}, where the latent space representation of the related ground state structures are dominated either by $\mathbf{v}_1$ or by combinations of $\mathbf{v}_1$ and $\mathbf{v}_2$, largely corresponds to the unclassified white region occurring in Fig.~\ref{fig:phasediagram_antlanger}. In an effort to obtain an even more profound insight into the so far hidden structures, we apply in a subsequent step a $k$-means clustering analysis of the data.

Before proceeding in this direction we note that our set of data has also been inspected with the ``t-stochastic neighbor embedding'' (t-SNE) \cite{Maaten2008} method, which represents another quite useful tool for visually representing a set of data from a high-dimensional feature space in a low-dimensional latent space, \ie, in a few embedding coordinates; for details, we refer the interested reader to Ref. \cite{Hartl2020Thesis}.

\paragraph{$k$-means clustering of structural data}

The preceding PCA provides already clear evidence that so far unexplored and unidentified ground state phases are hidden in the incredibly rich plethora of ordered bilayer structures in the asymmetric Wigner bilayer system. A step towards a more systematic analysis of the ground state configurations can be realized  by applying a subsequent $k$-means clustering analysis (see Subsec. \ref{subsubsec:k_means}) of the representation of the data set in terms of the nine leading principal components.

Before proceeding, three comments are in order: 

\begin{itemize}
\item[(i)] first we have to fix the actual parameter of the $k$-means clusterings, namely the number of clusters, $K$, which is not known {\it a priori};
\item [(ii)] when applying the $k$-means algorithm the choice of the initial location of the $K$ different clusters is usually arbitrary; however, the final result may depend on the particular choice of the initial cluster coordinates and on the initial assignment of the different data points to these clusters. It is therefore good practice to apply the $k$-means clustering several times with independent initial conditions. The results of these independent clusterings (termed $\mathbf{k}_1,\,\mathbf{k}_2,\dotsc$) can then be analysed, for instance in terms of adjusted mutual information, $I_K(\mathbf{k}_i,\mathbf{k}_j)$ (as defined and discussed in Appendix \ref{app:adjusted_mutual_information}); in our investigations, we have used 40 independent clusterings for a given $K$-value; the detailed results of these investigations are presented in Appendix \ref{app:reliability};
\item[(iii)] to simplify the analysis we have reduced the data set $\mathbf{X}^\mathrm{(asym)}$ by eliminating those trivial data points that are hexagonal monolayers and which are unambiguously characterized by $x = 0$. Ruling out these data points (which cover a large portion of the $(A, \eta)$-plane) leads to the data set  $\mathbf{X}^\mathrm{*}$ which covers thus all data points of the original data set except for those feature vectors of hexagonal monolayers. This elimination of data  reduces the size of data but does not have any influence on the PCA part of our approach, as shown by the results obtained for the explained variance PEV, $\lambda_i^{({\rm e})}$ (see Fig.~\ref{fig:wigner:phasediagram:clustering:pev}). Henceforward all quantities based on the reduced data set, $\mathbf{X}^\mathrm{*}$, are specified by an asterisk and we refer to the $k$-means clustering of the data set $\mathbf{X}^\mathrm{*}$ as $k^*$-means clustering.  
\end{itemize}

With the PCA results for the available structures at hand, we can now proceed to the obvious subsequent steps: we have to find an appropriate value for $K$ for the $k$-means classification of the structural data for which we have to analyze the results of several independent clusterings in order to identify an accurate labeling of the data.

In practice, we proceed as follows: first, we define a reasonable range of $K$ values, ranging in our case from $14$ to $42$ in integer steps. For each $K$-value, we then perform $40$ independent clusterings from which we choose the most suitable one; the rather lengthy procedure of how to compare the different clusterings and how to select the ``best'' labeling for a given $K$-value has been deferred to Appendix \ref{app:reliability}. This results in a set of ``best'' clusterings, one for each $K$-value. In an effort to identify the most appropriate number of clusters (or, in our case, the number of structural families), $K$, we compare the clusterings of this set of $K$-dependent ``best'' clusterings in an analogous procedure as described in Appendix \ref{app:reliability}: instead of comparing the results of independent clusterings for a given value of $K$, we now compare the ``best'' clusterings for different $K$-values (for more details also see Ref. \cite{Hartl2020Thesis}). The ``best'' clustering of the latter step is then labeled as  the ``best'' $k$-means (or $k^*$-means) clustering of the data set $\mathbf{X}$ (or $\mathbf{X}^\mathrm{*}$), and represents our revised mapping of the structural data set of Refs.~\cite{MoritzAntlanger2015,Antlanger2016,Antlanger2018} into families of structures.

\begin{figure}[htbp!]
\begin{center}
\includegraphics[width=0.8\textwidth,clip=True]{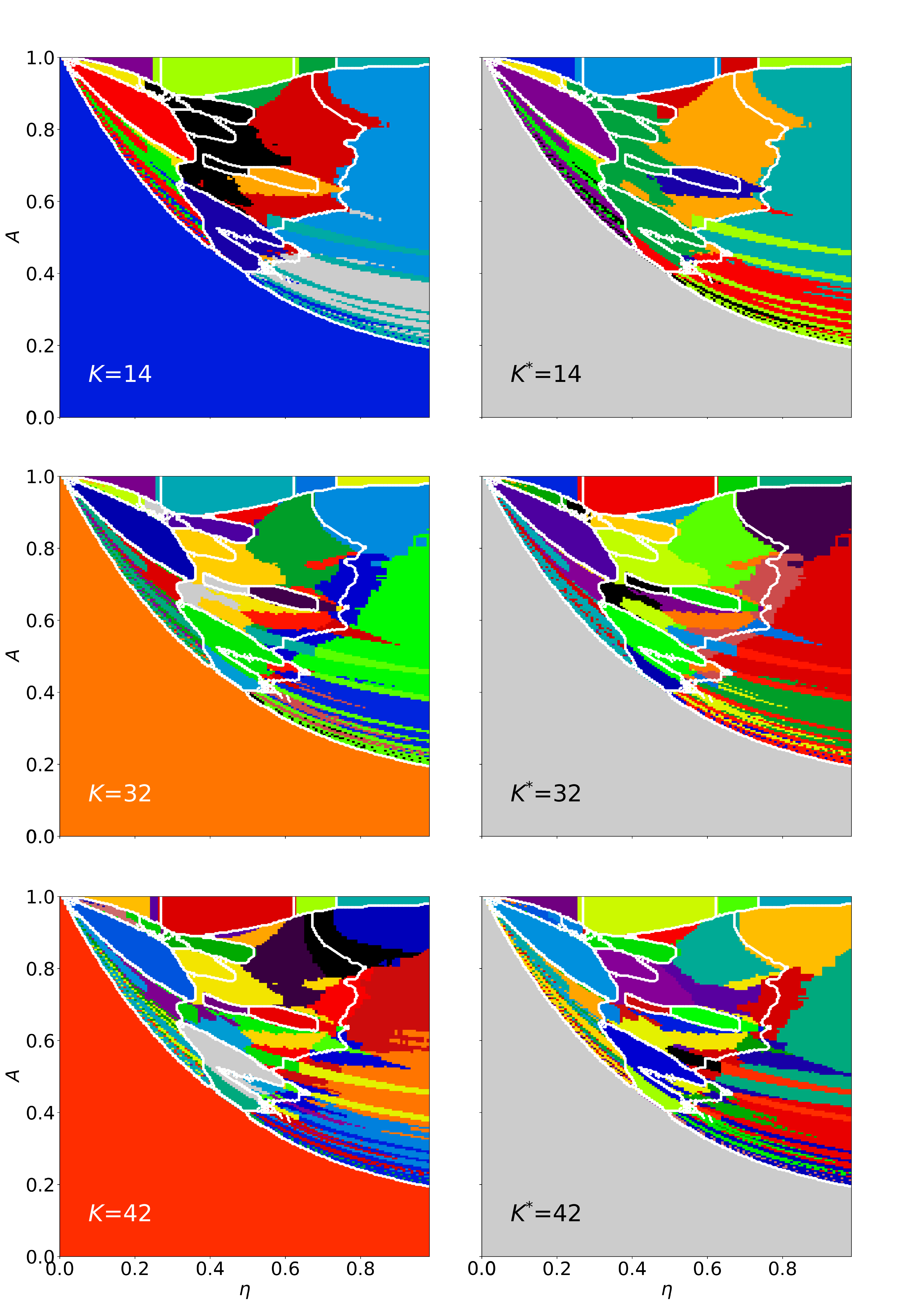}
\caption{Labeling the ground state configurations of  Refs.~\cite{MoritzAntlanger2015,Antlanger2016,Antlanger2018} into $K = 14$ (top row), $K = 32$ (middle row), and $K = 42$ (bottom row) families by $k$-means (left) and $k^*$-means (right) clusters. The color scheme is arbitrary, white lines indicate phase boundaries as specified in Refs.~\cite{MoritzAntlanger2015, Antlanger2016, Antlanger2018}.
While there are clearly differences in the clustering between the left and right columns for a given value of $K$ and $K^*$, respectively, we emphasize that these are not relevant for the discussion here.}
\label{fig:wigner:clustering}
\end{center}
\end{figure}

In an effort to visualize the impact of the value of $K$ on the $k$- (or $k^*$-)means results, we briefly summarize in the following the results obtained for three selected values of $K$, namely $K=$ 14, 32, and 42 (for a more detailed and graphical representation we refer to Ref. \cite{Hartl2020Thesis}). To this end we have redrawn in Fig.~\ref{fig:wigner:clustering} the diagram of states of our system for these three values of $K$, showing the respective phase labeling as suggested by the best clustering results of several $k$-means (left panels) and $k^*$-means (right panels) clustering procedures; note in this context that the color-coding of the different families is arbitrary. From the panels, it is obvious that the actual value of $K$ has a major impact on the final $k$-means (or $k^*$-means) results \footnote{We report at this occasion that for all $K$-values both $k$-means and $k^*$-means clustering results nicely correlate when evaluating the adjusted mutual information scores between the $k$-means and $k^*$-means clusterings, \ie, $I_K(\mathbf k_i, \mathbf k^*_j)$ as defined in the Appendix; for more details and graphical representations we refer to Ref. \cite{Hartl2020Thesis}}.

The value ${\bf K = 14}$ corresponds to the number of phases that have been identified in \cite{MoritzAntlanger2015, Antlanger2016, Antlanger2018} and that are specified in Table \ref{tbl:labels}; the corresponding clusterings are shown in the top row of Fig.~\ref{fig:wigner:clustering}. Already for $K=14$, most of the phases specified so far in literature \cite{MoritzAntlanger2015, Antlanger2016, Antlanger2018} are correctly identified. Phases \I~ through \Vs~ are clearly visible with (almost) correct boundaries; also the honey-comb phase \Honey, the phase \IIx, as well as phase \Pthree~ are identified essentially in a correct manner. Further, the phase boundary of the \Vx~ phase is resolved with good accuracy. However, a total number of $K$=14 (or $K^*$=14) clusters is definitely too small to resolve appropriately further details of the phases \SnubSquare, \SnubSquareTwo, \Pone, and \Ptwo; it also becomes obvious that the vast white regions of so far unclassified structures (see, e.g., Fig. \ref{fig:phasediagram_antlanger}) and/or the rich variety of yet unidentified substructures in phases \Ix~ and \Vx~  call for a closer and more refined analysis; thus a value $K = 14$ is definitely not appropriate to reach these goals. 

Proceeding to ${\bf K = 32}$ we find that -- and, admittedly, for the time being at the qualitative level -- our above requirements are met at a more satisfactory level. By a careful inspection of the related panels of Fig.~\ref{fig:wigner:clustering}, it seems that a set of 32 structural families is able to capture the variety of emerging structures: on one hand, this clustering does indeed accurately resolve the different phases identified in Refs.~\cite{MoritzAntlanger2015, Antlanger2016, Antlanger2018} (such as phases \I~ through \Vs~ as well as the phases \IIx, \Honey, \Pone, \Pthree~ and \SnubSquareTwo); on the other hand a value of $K = 32$ provides clear evidence of a rich variety of substructures in phases \Ix~ and \Vx~ that have not been classified so far in Refs.~\cite{MoritzAntlanger2015, Antlanger2016, Antlanger2018} and that are not captured by a $(K = 14)$-clustering, either. 

Eventually, a value of ${\bf K = 42}$ was -- from the conceptual point of view -- considered as the upper limit: increasing further the value of $K$ has led to the emergence of ``new'' phases which were -- after all -- only an artificial subdivision of well-defined phases. 
However, a closer comparison of the information contained for $(K = 42)$- and for $(K = 32)$-structure families -- via an analogous procedure as described in Appendix \ref{app:reliability} -- reveals, that the former one does not provide more substantial information on the emerging structure families than the latter one (we again refer to Ref. \cite{Hartl2020Thesis} for details). 

Thus, we eventually select the ``best'' $k^*$-means clustering (c.f., Appendix \ref{app:reliability}) for a total number of $(K = 32)$ clusters as the ``best'' labeling of the structural data set from literature \cite{MoritzAntlanger2015, Antlanger2016, Antlanger2018} of the asymmetric Wigner bilayer system into structure families. We henceforward refer to this result as \kmeans~ results and to the $c=(1,\dotsc,32)$ different clusters (\ie, to the different categories of structural families) as \kstar{c} families, respectively. 

\begin{figure}[h!tbp]
    \hspace{-0.05\textwidth}
    \includegraphics[width=1.0\textwidth, clip=True]{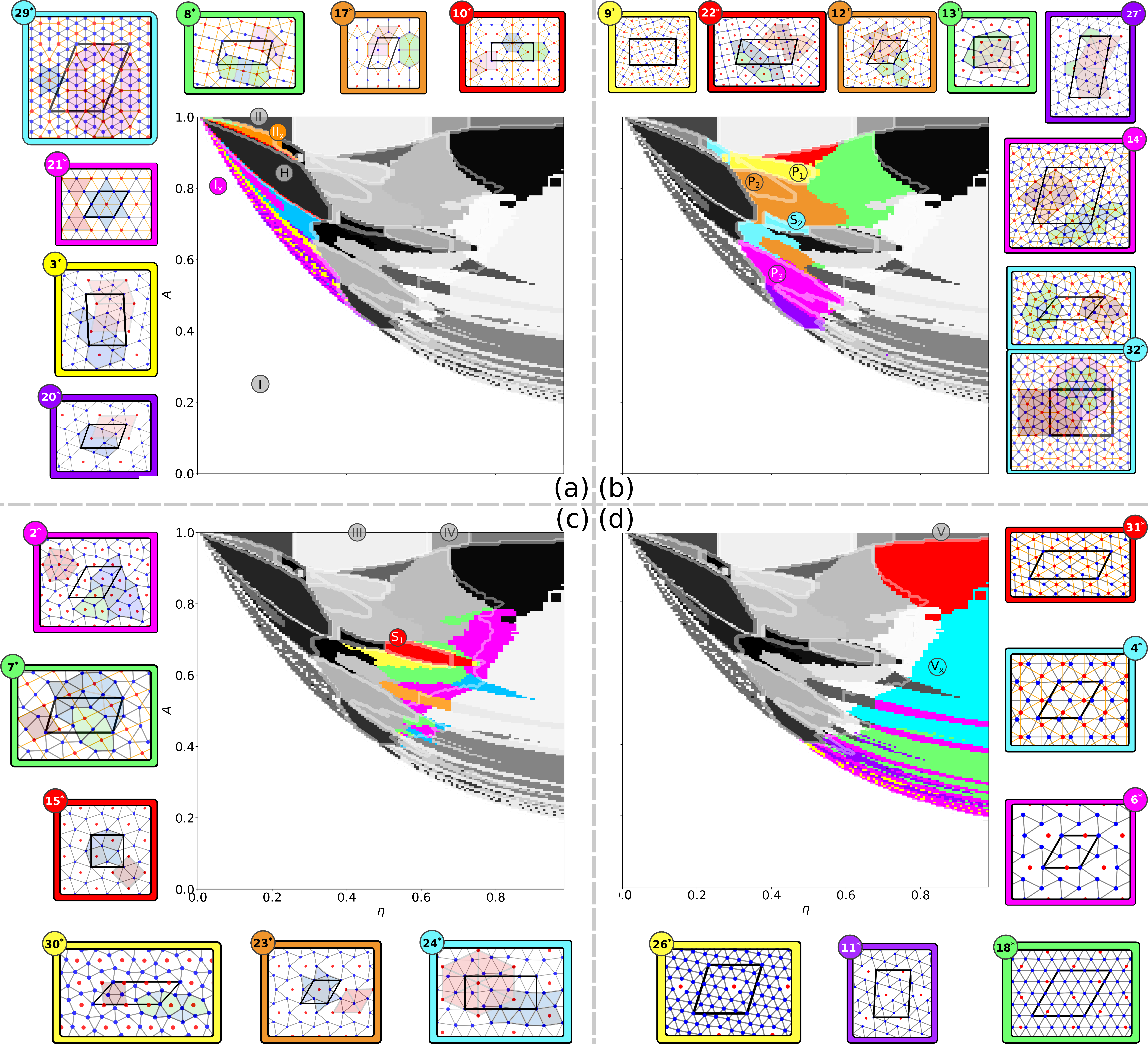}
    \caption{Ground state phase diagram of the asymmetric Wigner bilayer system (see Refs. \cite{MoritzAntlanger2015, Antlanger2016, Antlanger2018}) in the $(A, \eta)$-plane as identified by the $K^*$=32-means clustering algorithm; the respective 32 families of structures, \kstar{c}, are color-coded in different gray scales, ranging from white to black.
    The presentation of the ground state families of this phase diagram was subdivided into four $(A, \eta)$-subpanels (a)-(d) by symmetry arguments of the respective occurring ground states (see discussion of the subpanels in the text). In each subpanel we highlight the respective parameter regions in bright colors (cyan, yellow, green, orange, red, magenta, or purple -- in no particular order) where certain \kstar{c} families form the ground state; archetypical structures of the respective \kstar{c} families are shown as insets. In each panel these structures are labeled by the corresponding value of $c = 1, ..., 32)$ in the upper left corner to address their association to a certain family \kstar{c}.     
    For convenience, the frames of the insets are color-coded in the same way as the ground state regions of the respective \kstar{c} family in the phase diagram. Particles in ${\cal L}_1$ (${\cal L}_2$) are always colored blue (red) and connections between nearest neighbors in each layer are drawn. Special tiles and features of the different structures are highlighted by colored shapes and the respective unit cell of each structure is emphasized by a thick black frame. The phase-boundaries, as documented in literature \cite{MoritzAntlanger2015, Antlanger2016, Antlanger2018} (cf. also Fig.~\ref{fig:phasediagram_antlanger}) are indicated by opaque white lines in each panel; further, the corresponding phases (known from literature) are labeled by their acronyms in circles which are associated to the considered parameter region of the different panels. Colored, disk-shaped labels in subpanels (a) to (d) indicate, that archetypical structures of the corresponding \kstar{c} families are displayed as insets in the respective panels, while gray, disk-shaped labels (\ie~, \I~ through \Vs~ and \Honey) indicate, that the corresponding structures are not shown (although the phases have been identified by the \kmeans~ algorithm).
    }
    \label{fig:wigner:phasediagram:clustered:kmeans_prime_32}
\end{figure}

We first replace the original diagram of states (cf. Fig.~\ref{fig:phasediagram_antlanger}) with a diagram of states based on the \kmeans~ results, summarized in Fig.~\ref{fig:wigner:phasediagram:clustered:kmeans_prime_32}. Indeed, the 32 families of structures provide information about new, so far unidentified ground state configurations: we observe that some regions (such as the ones originally occupied by the \Ix~ or \Vx~ phases) are obviously subdivided into sub-regions, indicating the emergence of so far unclassified phases. For each of these families (and except for the ``trivial'' phases \I~ through \Vs~ and \Honey) we present in Fig.~\ref{fig:wigner:phasediagram:clustered:kmeans_prime_32} typical examples of ground state configurations of the system for different values of $A$ and $\eta$, based on the \kstar{c}  clustering. In an effort to obtain a better overview of the different structures, we have split in Fig.~\ref{fig:wigner:phasediagram:clustered:kmeans_prime_32} the presentation of the diagram of states into four qualitatively different subpanels (labeled (a) to (d)):

\begin{itemize}
\item
in {\bf panel (a)} we focus on the region which hosts the phases \Ix, \II, \IIx~ and \Honey~, occurring at small to medium values of $\eta$ (\ie, $0<\eta\lesssim0.4$) and medium to large values of $A$ (\ie, $0.4\lesssim A\leq1$); the ground state structures in this region have in common that they form a hexagonal monolayer if all particles were projected onto the same layer;
\item in {\bf panel (b)} we collect structural families which feature pentagonal tiles in ${\cal L}_1$, \ie, configurations that belong to the broader family of pentagonal structures: suggested ground states candidates that belong to this category are \Pone, \Ptwo, \Pthree~ and \SnubSquareTwo, the associated range of the system parameters can roughly be given by $0.3\lesssim\eta\lesssim0.7$ and $0.4\lesssim A\lesssim 0.9$; 
\item in {\bf panel (c)} we address \kstar{c} families which have tilings in ${\cal L}_1$ that are similar to the snub-square structure, \SnubSquare, which, in turn, can potentially give rise to ground state configurations with global twelvefold symmetry~(see Refs. \cite{Stampfli1986, Oxborrow1993, Grunbaum1987});
\item eventually, in {\bf panel (d)} we present \kmeans~ results which can be related to the \Vx~ region in the parameter space of the asymmetric Wigner bilayer system, \ie, at large plate separation distances ($\eta \gtrsim 0.7$) and covering a large range of $A$. 
\end{itemize}

These results will be discussed in detail in the following Subsection.

\subsection{New insights to the Wigner bilayer system from unsupervised learning}
\label{subsec:new_insights}

A comparison of the identified phases of the asymmetric Wigner bilayer system via methods from literature \cite{MoritzAntlanger2015, Antlanger2016, Antlanger2018} (see Fig.~\ref{fig:phasediagram_antlanger}) and via labeling by unsupervised \kmeans~ techniques (see Fig.~\ref{fig:wigner:clustering}) shows -- on one hand -- an excellent agreement for several structural families: not only the symmetric cases (\ie,  phases \I~ through \Vs), but also more complex configurations such as \IIx, \Honey, \SnubSquare, \SnubSquareTwo, \Pone, \Ptwo, and \Pthree~ are faithfully reproduced. However, the clustering technique is able to reveal that for several regions in the parameter space (such as the regions of the phases \Ix~ and \Vx) a re-analysis of the data is in order; this will be done in the following.

\begin{itemize}
\item[(a)]  {\bf phase \Ix}: based on the clustering (see Fig.~\ref{fig:wigner:phasediagram:clustered:kmeans_prime_32}), we can indeed identify a rich variety of phases within this region, which is correlated with the dodecagonal $\Psi_{12}^{(4)}$ ``hole'' BOOP; note in this context, in particular the blue region within the \Ix~ phase near the boundary to the  \Honey~phase and the relatively large region populated by the \kstar{29} family (see panel (a) of Fig.~\ref{fig:wigner:phasediagram:clustered:kmeans_prime_32}, marked in cyan); 
\item[(b)] {\bf pentagonal phases}: regions that were declared in the original diagram of states to be populated exclusively by the \Pthree~ structure (see Fig.~\ref{fig:phasediagram_antlanger}) now show -- as a consequence of the more refined analysis of the clustering approach -- a considerably more complex internal structure: we can see from panel (b) of Fig.~\ref{fig:wigner:phasediagram:clustered:kmeans_prime_32} that not only the \kstar{14} and the \kstar{27} families, but also the \SnubSquareTwo~ configuration (represented by the \kstar{32} family) form similar patterns in ${\cal L}_2$ (notably distorted rectangles and triangles organized in a distorted snub-square vertex); however, the decoration of the ${\cal L}_2$ tiles with particles in ${\cal L}_1$ becomes increasingly complex: more particles of ${\cal L}_1$ are involved per tile in ${\cal L}_2$ the more we approach the boundary to phase \I~ in the parameter space.
The structure in ${\cal L}_1$ thus approximates a trigonal lattice when approaching the phase boundary of the \I~ phase, and only exhibits pentagonal defects around the $x,y$-locations of ${\cal L}_2$ particles.
Interestingly, the structural families \kstar{27} and \kstar{14} appear to belong to a family of increasingly complex super-structures of the top panel of family \kstar{32}, \ie, the original \SnubSquareTwo phase.
This might hint at the existence of a low energy quasi-crystalline structure of this family in the asymmetric Wigner bilayer system.
\item[(c)] {\bf phases \SnubSquare~ and \SnubSquareTwo}: with the help of the clustering analysis we are now able to classify also the previously unclassified structures in the $(A, \eta)$-regions in the vicinity of phases \SnubSquare~ and \SnubSquareTwo~ (cf. related white regions in Fig.~\ref{fig:phasediagram_antlanger}) by several different structural families, as illustrated in panel (c) of Fig.~\ref{fig:wigner:phasediagram:clustered:kmeans_prime_32}:  these structures have in common that their basic tiles (such as equilateral triangles and squares arranged in a snub-square vertex), form in ${\cal L}_1$ a structure that might indicate the existence of a quasicrystalline state with global dodecagonal symmetry~\cite{Stampfli1986, Oxborrow1993, Janssen2007}. 
\item[(d)] {\bf phase \Vx}: eventually, in the \Vx~ region of the diagram of states some interesting new structural families are identified as a consequence of the \kmeans~ procedure, as can be seen in panel (d) of  Fig.~\ref{fig:wigner:phasediagram:clustered:kmeans_prime_32}. Characteristic values and boundaries of the order parameters and of the corresponding principal component representation of all newly identified structural families \kstar{c} depicted in Fig.~\ref{fig:wigner:phasediagram:clustered:kmeans_prime_32} are collected in Subsection 3.1.7 of \cite{Hartl2020Thesis} (which also provides more detailed information about the symmetries of these structural families).
\end{itemize}


                        
\section{Conclusions} 
\label{sec:conclusions}

In this contribution, we have re-analyzed the ordered ground state configurations of the asymmetric Wigner bilayer system where identical point charges are immersed into the space confined between two parallel plates of opposite charge. The ratio of the (not necessarily equal) surface charges of the two plates ($\sigma_1$ and $\sigma_2$), \ie, $A = \sigma_2/\sigma_1$ and the reduced, dimensionless distance $\eta$ between the plates uniquely define each state point of the system. A previous classification scheme of the emerging configurations \cite{MoritzAntlanger2015,Antlanger2016,Antlanger2018} into structural families, was done by ``hand'': such an approach is not only a tedious, possibly hopeless task, but it is also -- and even more relevant -- prone to faulty analysis, preventing thereby a faithful identification and sorting of the emerging structures. 

In an effort to overcome these drawbacks and imponderabilities we have re-analyzed this huge set of data (comprising $\sim$ 60~000 data points) and have used instead machine learning based tools, notably on the principal component analysis and a subsequent $k$-means clustering algorithm. In a first step, we have assigned to each  emerging structure a feature vector with 30 entries: (i) predominantly suitably defined order parameters that characterize the ground state configuration, (ii) the composition of the system, and (iii) further structural information based on the radial distribution function. With the help of a principal component analysis we have extracted from this representation the most relevant information by projecting the underlying 30-dimensional feature space on a reduced representation in the so-called latent space; in our case, we found out that a dimension of nine of the latent space is sufficient to capture the relevant features. Eventually, we have classified this reduced information via a $k$-means algorithm and have collected ground state configurations into families of structures that occupy ``neighboring'' regions in this latent space. Applying different types of thorough internal consistency checks we eventually found that the sorting of the emerging structures into 32 families provides the most reliable and most consistent classification scheme of these structures, as compared to 14 structural families that were identified in the preceding, ``by hand''  classification approach \cite{MoritzAntlanger2015,Antlanger2016,Antlanger2018}. 

In view of the achieved results presented in this contribution, we can righteously conclude that with our machine learning based tool at hand we are now able to provide a systematic, reliable, and thorough classification scheme for the emerging structures. The new insights into the diagram of states comprise now particle configurations that were previously hidden (or even not accessible) in the zoo of structures in the database of Refs.~\cite{MoritzAntlanger2015,Antlanger2016,Antlanger2018}: (i) within the region that was originally assigned to the \Ix~ structure we could identify a rich variety of phases which are characterized by a dodecagonal $\Psi_{12}^{(4)}$ bond order parameter; (ii) the region that was originally thought to be populated exclusively by the \Pthree structure has a very rich internal structure and other configurations could be identified that are formed by distorted rectangles and triangles, organized in a distorted snub-square vertex; (iii) it could be shown that the previously unclassified (``white'') regions in the vicinity of phases \SnubSquare and \SnubSquareTwo are populated by several different structural families; all these structures have in common that their basic tiles arrange in ${\cal L}_1$ into a structure that might possibly indicate the existence of a quasicrystalline state with global dodecagonal symmetry; (iv) eventually, the huge, and previously unexplored region populated by the \Vx~ structure reveals the existence of quite a few interesting subtle new structural families. With all these new findings we conclude that the re-analyzed diagram of states is bare of any white (\ie, unexplored) regions.

Apart from a more systematic and thorough classification of the structures, our approach offers several other attractive features that turned out to be very useful. 

A particular challenge when identifying the ground state configurations for our system is the issue of degeneracy: this applies to structures that were obtained in Refs.~\cite{MoritzAntlanger2015,Antlanger2016,Antlanger2018} via 
evolutionary algorithms, involving different numbers of particles per unit cell but characterized by the same composition; they eventually end up as identical structures with the same (\ie, degenerate) energies but parameterized by different (but equivalent) unit cells. In our previous ``by hand'' classification scheme it was very difficult to prove the structural equivalence of two such particle configurations, while with the clustering-based labeling of the available structural database as it has been used in this contribution such degenerate configurations are automatically grouped into the same structural family via the information emerging from the feature vectors. 

Another advantage of our approach is that we easily obtain within each structural family an energy-based ranking of the structures: thus for a given state point (defined by a pair $(A, \eta)$) we have clear information about an energy-based ranking of the structures, starting at the lowest level with the ground state configuration. In this manner we can identify those structures that energy-wise are sometimes very close to the ground state configuration, but which possibly are structure-wise distinctively different from the latter one;
such a ranking was essentially inaccessible in our previous approach while in our present approach they are automatically labeled. With this information at hand, we can then focus in a subsequent step on different structural families, where we have direct access to the properties of the energetically competing families of structures for any $(A, \eta)$-state (see a more in-depth discussion in Appendix \ref{app:degeneracy}). 

From a more formal point of view, it should be mentioned that, in general, clustering of structural data using PCA and $k$-means clustering (or any other, suited clustering algorithm or classification algorithm) provides us with an additional attractive feature: PCA is a linear transformation from the feature space to the latent space and $k$-means is a mapping of a data point in the latent space to a cluster label. Once the clustering algorithm is trained (\ie, once it has converged) it can be used as a classification model \footnote{Nowadays it is easily possible to train a neural network in a supervised way with the objective of performing classification tasks \cite{HIGHBIAS2019}. For our purposes, such a task would be to classify structural data into a number of $K$ different categories (identified, for instance, by unsupervised clustering) which would allow us to directly classify a structure from its geometric, structural data, e.g. via coordinates and lattice vectors~\cite{Guttenberg2016}. The output of the classifier would then be the probability of a structure falling into any of the $K$ clusters or families (when using ``softmax'' activation in the output layer of the neural network and ``categorical cross-entropy loss'' during training \cite{HIGHBIAS2019}), which may give additional insight when comparing competing structures.}  and we can ask the following questions for an arbitrary structure: ``what family would it belong to?'', ``where would it appear in the phase-diagram'' and ``what would be its characteristic features?''; see the Appendix A.1.2 of \cite{Hartl2020Thesis} for related numerical details on the characteristic features of the here employed \kmeans~ classification scheme of the structural data from Refs.~\cite{MoritzAntlanger2015,Antlanger2016,Antlanger2018}.

When re-exploring the data set of configurations of the asymmetric Wigner bilayer system we also encountered situations where particular numerical care had to be taken: this applies in particular when exploring regions where first-order transitions between competing structures have to be identified, characterized by distinct discontinuous changes in order parameters. In an effort to locate the transition point accurately, particular numerical care has to be taken. Even more challenging are second-order phase transitions (such as those between phases \II $\rightarrow$ \III~ and \III $\rightarrow$ \IV) where some of the features (order parameters, etc.) change continuously.

Summarizing, we can righteously state that the clustering tools discussed in this contribution represent an indispensable help in classifying complex emerging structures and undoubtedly offer a deeper insight into the complexity of the phase diagram of the asymmetric Wigner bilayer system.


\section*{Acknowledgements}

We acknowledge Moritz Antlanger for helpful discussions.
BH acknowledges funding via a DOC fellowship during the time of his Ph.D. which made the presented research possible.
MM was also supported by Slovak Grant Agency VEGA No. 2/0144/21.

\bibliography{main}

\eject

\begin{appendix}



\section{Adjusted mutual information}
\label{app:adjusted_mutual_information}

In the following we provide a tool that is able to test the reliability of the partitioning of a data set, $\mathbf{X}=\{\mathbf{x}_1,\dotsc,\mathbf{x}_N\}$, of $N$ data elements, into subsets, $\mathbf{U}^R = \{\mathbf{U}_1,\mathbf{U}_2,\dotsc,\mathbf{U}_R\}$, with the following requirements:  $\cup_{i=1}^R \mathbf{U}_i= \mathbf{X}$ and $\mathbf{U}_i \cap \mathbf{U}_j = \emptyset$ for all $i\neq j$ \cite{Vinh2010}.

Commonly used clustering algorithms, such as $k$-means clustering or DBSCAN \cite{Ester1996} are, on the one hand, applicable in a variety of problems, but, on the other hand, not unique in their predictions: the final result of such algorithms usually depends on the clustering algorithm (such as the initial -- usually random -- choice of assigning data points to clusters, etc.), on the choice of the parameters of the algorithm, or on noise in the data~\cite{SKLEARN}.

It is thus of particular relevance to comparing the results of different clusterings of a given data set ${\bf X}$. Thus we want either (i) to compare the results emerging from different clustering algorithms or (ii) to compare the results of the same algorithm but with different initial conditions. Assuming two different partitionings of a data set ${\bf X}$, \ie ${\bf U} \equiv \mathbf{U}^R = \{\mathbf{U}_1,\mathbf{U}_2,\dotsc,\mathbf{U}_R\}$ and ${\bf V} \equiv \mathbf{V}^C = \{\mathbf{V}_1,\mathbf{V}_2,\dotsc,\mathbf{V}_C\}$ (satisfying both the above requirements), we want to quantify their overlap, or, in other words, quantify the shared information of the two different clusterings.

A fundamental class of techniques for comparing clusterings of labeled data sets is based on  information theoretic measures \cite{Vinh2010}. In our contribution, we use the concept of adjusted mutual information \cite{Vinh2009, Vinh2010}.

In a first step we define the $(R\times C)$-dimensional contingency table $M=[n_{ij}]_{j=1\dotsc C}^{i=1\dotsc R}$ (see Table \ref{tab:methods:order:contingency_table}), whose elements, $n_{ij}=|\mathbf{U}_i \cap \mathbf{V}_j|$, quantify the number of common objects in $\mathbf{U}_i$ and $\mathbf{V}_j$. The mutual information, $I_M(\mathbf{U}, \mathbf{V})$, of two different clusterings, $\mathbf{U}$ and $\mathbf{V}$, is defined as~\cite{Vinh2009,Vinh2010}

\begin{equation}
  I_\mathrm{M}(\mathbf U, \mathbf V) = \sum_{i=1}^R\sum_{j=1}^C P_\mathbf{UV}(i, j)\log\frac{P_\mathbf{UV}(i, j)}{P_\mathbf U(i)P_\mathbf V(j)}
\end{equation}
where $P_\mathbf {UV}(i, j)=|\mathbf{U}_i\cap \mathbf{V}_j|/N$ is the probability that a (random) data point belongs to both clusters $\mathbf{U}_i$ (in $\mathbf U$) and $\mathbf{V}_j$ (in $\mathbf V$); $P_\mathbf U(i)=|\mathbf{U}_i|/N$ and $P_\mathbf V(j)=|\mathbf{V}_j|/N$ denote the probabilities that randomly chosen data points fall into the cluster $\mathbf{U}_i$ and $\mathbf{V}_j$, respectively.
In that way, $I_\mathrm{M}(\mathbf U, \mathbf V)$ quantifies the information which is shared between two clusterings and thus can be interpreted as a similarity measure for clusterings; notably, the upper bounds of $I_\mathrm{M}(\mathbf U, \mathbf V)$ are the quantities $H(\mathbf U)=-\sum_{i=1}^R P_\mathbf U(i)\log P_\mathbf U(i)$ and $H(\mathbf V)=-\sum_{j=1}^C P_\mathbf V(j)\log P_\mathbf V(j)$ \cite{Vinh2010}.

\begin{table}[h!t]
  \centering
  \begin{tabular}{c|cccc|c}
  $\mathbf{U}^R / \mathbf{V}^C$ & $\mathbf{V}_1$ & $\mathbf{V}_2$ & $\dotsc$ & $\mathbf{V}_C$ & Sums \\ \hline
  $\mathbf{U}_1$ & $n_{11}$ & $n_{12}$ & $\dotsc$ & $n_{1C}$ & $a_{1}$\\
  $\mathbf{U}_2$ & $n_{21}$ & $n_{22}$ & $\dotsc$ & $n_{2C}$ & $a_{2}$\\
  $\vdots$  & $\vdots$ & $\vdots$ & $\ddots$ & $\vdots$ & $\vdots$ \\
  $\mathbf{U}_R$ & $n_{R1}$ & $n_{R2}$ & $\dotsc$ & $n_{RC}$ & $a_{R}$\\ \hline
  Sums  & $b_1$ & $b_2$ & $\dotsc$ & $b_C$ & $\sum_{ij} n_{ij} = N$ \\
  \end{tabular}
  \caption{Contingency table between two different clusterings, $\mathbf{U}^R = \{\mathbf{U}_1,\mathbf{U}_2,\dotsc,\mathbf{U}_R\}$ and $\mathbf{V}^C = \{\mathbf{V}_1,\mathbf{V}_2,\dotsc,\mathbf{V}_C\}$, with $n_{ij}=|\mathbf{U}_i \cap \mathbf{V}_j|$ being the number of common objects in clusterings $\mathbf{U}_i$ and $\mathbf{V}_j$; further $a_i=\sum_{j=1}^C n_{ij}$ and $b_j=\sum_{i=1}^R n_{ij}$.}
  \label{tab:methods:order:contingency_table}
\end{table}

The adjusted mutual information, $I_K(\mathbf{U},\mathbf{V})$, corrects the information-theoretic measures of mutual information agreement of clusterings for chance (see Refs.~\cite{Rand1971, Vinh2009, Vinh2010} for details),
and can be given by

\begin{equation}
  I_K(\mathbf U, \mathbf V)=\frac{I_\mathrm{M}(\mathbf U, \mathbf V) - E_\mathrm{MI}(\mathbf U, \mathbf V)}{\max[H(\mathbf U), H(\mathbf V)] - E_\mathrm{MI}(\mathbf U, \mathbf V) },
  \label{eq:methods:order:mutual_information}
\end{equation}
where the expected mutual information, $E_\mathrm{MI}(\mathbf U, \mathbf V)$, between two (random) clusterings is defined by

\begin{multline}
  E_\mathrm{MI}(\mathbf U, \mathbf V) =
  \sum_{i=1}^R\sum_{j=1}^C \sum_{n_{ij}=\max(1,a_i+b_j-N)}^{\min(a_i,b_j)}\frac{n_{ij}}{N}\log\left(\frac{N\,n_{ij}}{a_i b_j}\right)
  \times
  \\
  \frac{a_i\,!b_j\,!(N-a_i)\,!(N-b_j)\,!}{N\,!n_{ij}\,!(a_i-n_{ij})\,!(b_j-n_{ij})\,!(N-a_i-b_j+n_{ij})\,!},
\end{multline}
with $a_i=\sum_{j=1}^C n_{ij}$ and $b_j=\sum_{i=1}^R n_{ij}$ being the partial sums over the contingency table $M[n_{ij}]_{j=1\dotsc C}^{i=1\dotsc R}$ defined in Table \ref{tab:methods:order:contingency_table}.

A value of $I_K(\mathbf{U},\mathbf{V})=1$ indicates perfect overlap between two different clusterings (\ie, the two clusterings label the data equivalently but potentially use different numerical values to label the different clusters), a value smaller than one indicates differences in the clusterings.



\section{Analyzing phase-boundaries by comparing energetically degenerate but geometrically different families of structures throughout the phase diagram
}
\label{app:degeneracy}

In this Appendix, we discuss in more detail and on a more quantitative level how our combined approach (of a PCA and a subsequent $k$-means clustering) can cope with the issue of possibly degenerate structures emerging in the database of structures of our Wigner bilayer system. 

As noted in the body of the text, the main challenge is to identify for a given pair of $(A, \eta)$-values the energetically most favorable configuration: if such an identification is made ``by hand'' the following implications have to be expected: (i) extremely small energy differences between competing structures might occur and (ii) the fact that the genetic algorithm produces energetically degenerate structures, which are characterized by different unit cells that describe an equivalent lattice. These implications are discussed in the following.

\begin{figure}[htb!]
  \centering
    \includegraphics[width=\columnwidth, clip=True]{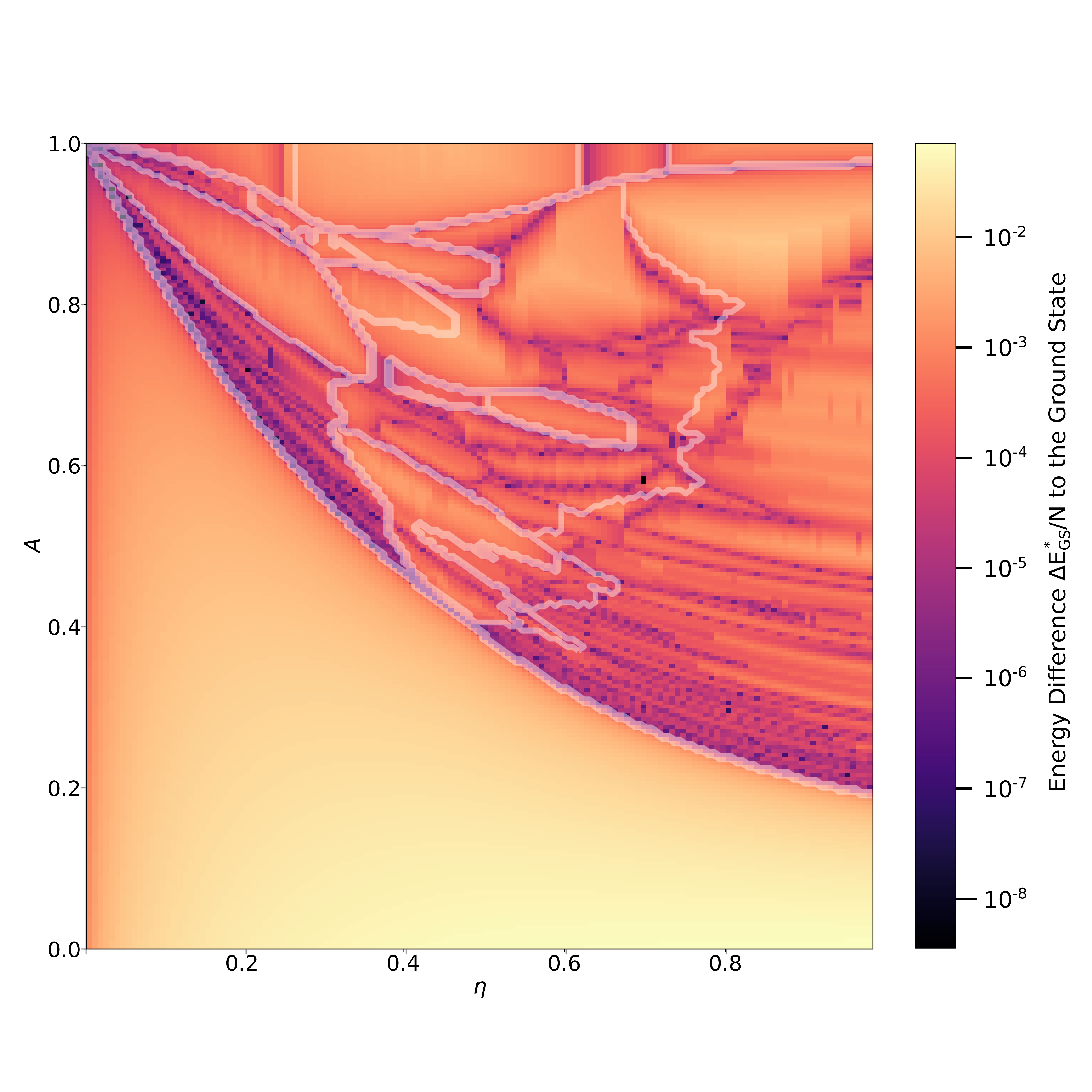}
   \caption{Difference, $\Delta E^*_\mathrm{GS}/N = (E^*/N - E^*_\mathrm{GS}/N)$, between the energy (per particle), $E^*/N$, of the energetically most favorable structure amongst all non-ground state structural families with respect to the respective ground state energy (per particle), $E^*_\mathrm{GS}/N$, of the asymmetric Wigner bilayer system for the data set taken from Refs.~\cite{MoritzAntlanger2015, Antlanger2016, Antlanger2018} for every state point in the $(A, \eta)$-plane. Opaque white lines indicate phase boundaries as presented in literature (see also Fig.~\ref{fig:phasediagram_antlanger}). In the proximity of the boundaries of the \Honey~ and \Ix~ phases and within the \Ix~ region we observe very small values of $\Delta E^*_\mathrm{GS}/N$ (ranging typically from $\approx 10^{-8}$ to $10^{-6}$), corresponding to nearly degenerate competing structures of different structural families; the related structures exhibit large values of the twelvefold symmetric order parameter $\Psi_{12}^{(4)}$~\cite{Antlanger2016, Hartl2020Thesis}.
   This region corresponds to the newly identified ground state candidate family \kstar{29} illustrated by the top left inset structure and in the cyan-emphasized area in the $(\eta,A)$-plane of Fig.~\ref{fig:wigner:phasediagram:clustered:kmeans_prime_32}(a); details of this family of structures are summarized in Sec.~\ref{subsec:new_insights} and will be discussed in more detail in a forthcoming contribution.}
    \label{fig:wigner:clustering:tie-line}
\end{figure}

The clustering algorithm helps us to classify at each $(A, \eta$)-point {\it all} the structures provided by the evolutionary algorithm within a certain  number of  structure families (in our case -- and as argued in the body of the text -- we have chosen 32 families). In Fig.~\ref{fig:wigner:clustering:tie-line} we display in a color-coded manner the energy difference between the energetically most favorable structure (\ie, the energy of the ground state, termed $E_{\rm GS}^*$) and the structure that pertains to the structure family with the energetically second best structure \footnote{Although most of the suggested ground state candidates identified by the evolutionary algorithm in Refs.~\cite{MoritzAntlanger2015, Antlanger2016, Antlanger2018} are very likely to represent the true ground state configurations of the asymmetric Wigner bilayer system at a given state point, there is no rigorous proof that they are, indeed, the ground states. Thus, whenever we use the term ``ground state solutions'' we refer to ``ground state \textit{candidate} solutions'' of the asymmetric Wigner bilayer system.}. Thus, dark-colored areas (notably in black and purple) in Fig.~\ref{fig:wigner:clustering:tie-line} highlight regions in the diagram of states where the energetically best and second best structures exhibit very small differences in their energies (going down to values as small as 10$^{-8}$ in relative units); note that this feature is particularly pronounced at phase boundaries. In contrast, orange to yellow areas in the $(A, \eta)$-plane indicate a large energetic gap between the ground state and energetically subsequent, competing structure.

\begin{figure}[htb]
\begin{center}
 \includegraphics[width=0.9\textwidth, clip=True]{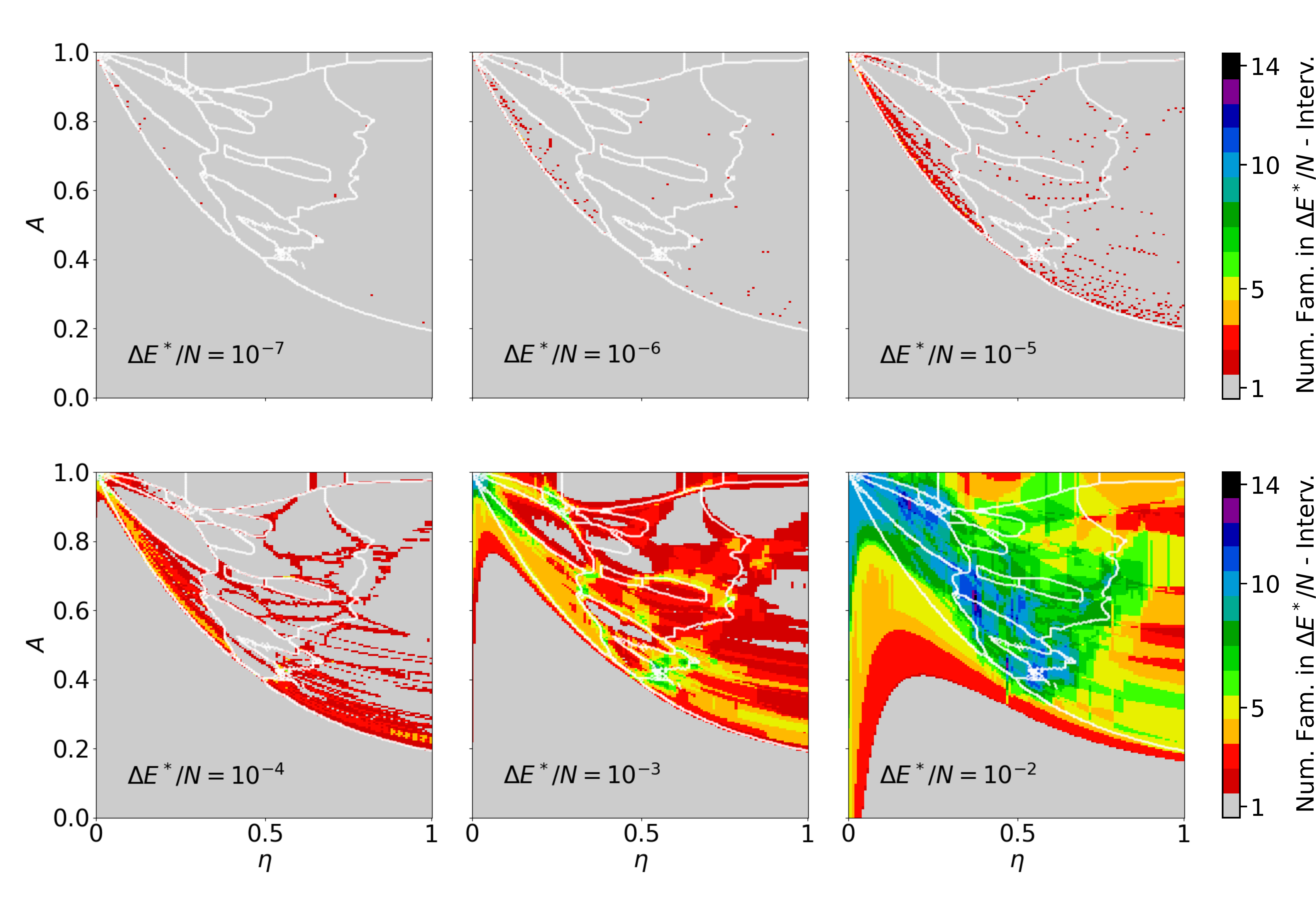}
\caption{Color-coded plot (cf. color bars on the right-hand side) of the numbers of 
    families ($K = 32$) that show an energy difference of at most $\Delta E^*/N$ with respect to the respective ground state candidates of the asymmetric Wigner bilayer system (suggested by Refs.~\cite{MoritzAntlanger2015, Antlanger2016, Antlanger2018}) in the $(A, \eta)$-plane. The values of $\Delta E^*/N$, range from $10^{-7}$ to $10^{-2}$, and are chosen separately for each panel (as labeled). Light-gray regions visualize regions where only one structure is identified within the respective $\Delta E^*/N$-interval; the colors according to the color bar emphasize the level of ``$\Delta E^*/N$-degeneracy'' at a given pair of the system parameters, \ie, the number of \kstar{c} families which exhibit an energy difference to the ground state -- at a given $(\eta,A)$-pair -- of at most $\Delta E^*/N$.
    Phase boundaries from Refs.~\cite{MoritzAntlanger2015,Antlanger2016,Antlanger2018} (cf. Fig.~\ref{fig:phasediagram_antlanger}) are emphasized by white lines.}
\label{fig:wigner:clustering:tie-line-competitors}
\end{center}
\end{figure}

An alternative view on the dataset provided in literature \cite{MoritzAntlanger2015, Antlanger2016, Antlanger2018} via the structural families (as obtained via the clustering algorithm) is shown in Fig.~\ref{fig:wigner:clustering:tie-line-competitors}: here we display -- again via a color code -- the number of \kstar{c} families whose energetically most favorable structure lies within an energy interval $\Delta E^*/N$ above the energy of the respective ground state, \ie, $E^*_\mathrm{GS}/N$. Assuming different values of $\Delta E^*/N$,  ranging in relative units from 10$^{-5}$ down to $10^{-7}$ (as labeled) to the ground state energy, some two or three structures of competing families have been identified with the genetic algorithm. These findings indicate, in turn, the high numerical accuracy that is required to distinguish between energetically competing structures (see also discussion in \cite{MoritzAntlanger2015, Antlanger2016, Antlanger2018}).

\section{
On the reliability of the clustering algorithm
}
\label{app:reliability}

\begin{figure}[htb]
\begin{center}
    \includegraphics[width=0.49\textwidth, clip=True]{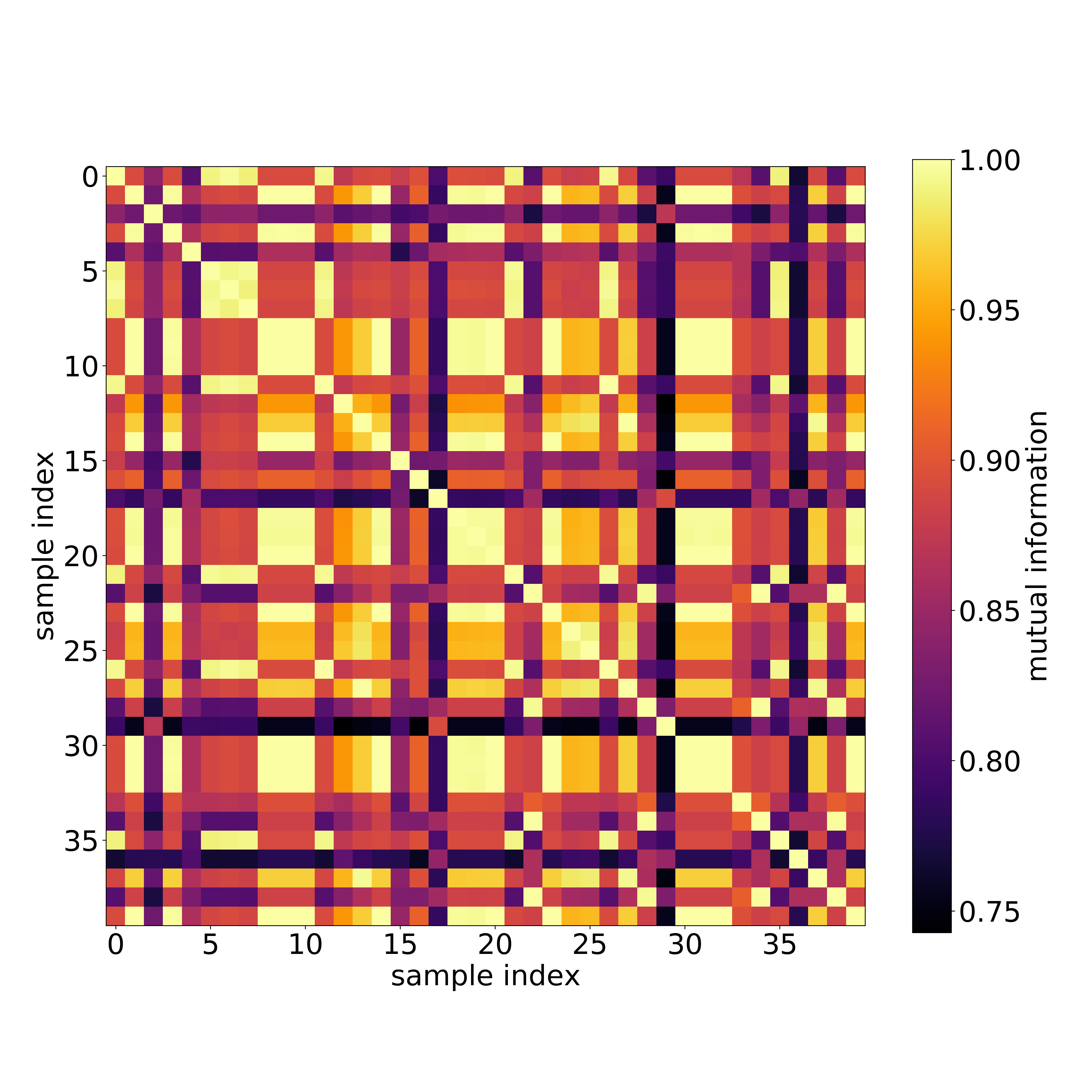}
    \includegraphics[width=0.49\textwidth, clip=True]{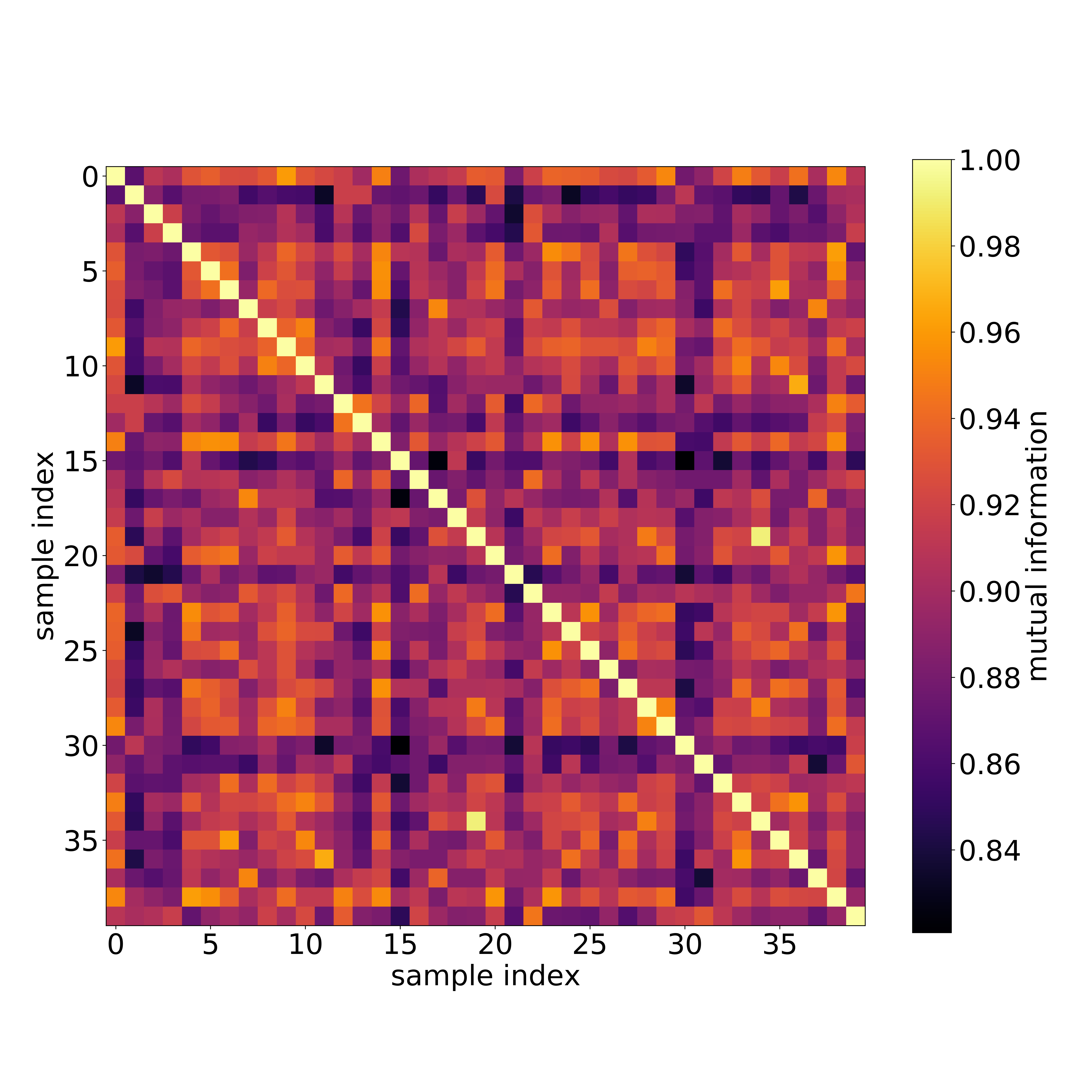}
    \caption{Adjusted mutual information score, $I_K(\mathbf k_i^*, \mathbf k_j^*)$, as defined by in Eq. (\ref{eq:methods:order:mutual_information}) of 40 different and randomly initialized $k^*$-means clustering results with $K=14$ (left) and $K=32$ (right) clusters, respectively. Values of $I(\mathbf k_i^*, \mathbf k_j^*)$ close to unity (bright, yellow regions) identify large overlap between the different clusterings, $\mathbf{k}^*_i$ and $\mathbf{k}^*_j$, while smaller values, \ie~ $I(\mathbf k_i^*, \mathbf k_j^*)\approx0.8$ (black and purple) indicate less consistent results. Note that $I_K(\mathbf k^*_i, \mathbf k^*_j)=I_K(\mathbf k^*_j, \mathbf k^*_i)$.}
\label{fig:wigner:phasediagram:clustering:kmeans:mutual}
\end{center}
\end{figure}

In Fig. \ref{fig:wigner:phasediagram:clustering:kmeans:mutual} we present the adjusted mutual information, $I_K(\mathbf{k}^*_i,\mathbf{k}^*_j)$, of $N_c=40$ independent clustering results (with $i,j=0,\dotsc,N_c-1$) of the $k^*$-means clustering algorithm for $K=14$ and $K=32$ clusters, respectively. 

For a smaller number of clusters (\ie~ $K=14$) the algorithm is more stable: many samples exhibit a perfect score of the adjusted mutual information, \ie~ $I_K(\mathbf{k}^*_i,\mathbf{k}^*_j) \simeq 1$ (cf. yellow pixels in the left panel of Fig. \ref{fig:wigner:phasediagram:clustering:kmeans:mutual}), indicating that the algorithm has identified the same results several times. For a larger number of clusters (\ie~ $K=32$), the situation is more complicated since the number of possible clustering results grows rapidly with the number of clusters. 

\begin{figure}[ht!]
\begin{center}
  \includegraphics[width=0.49\textwidth, clip=True]{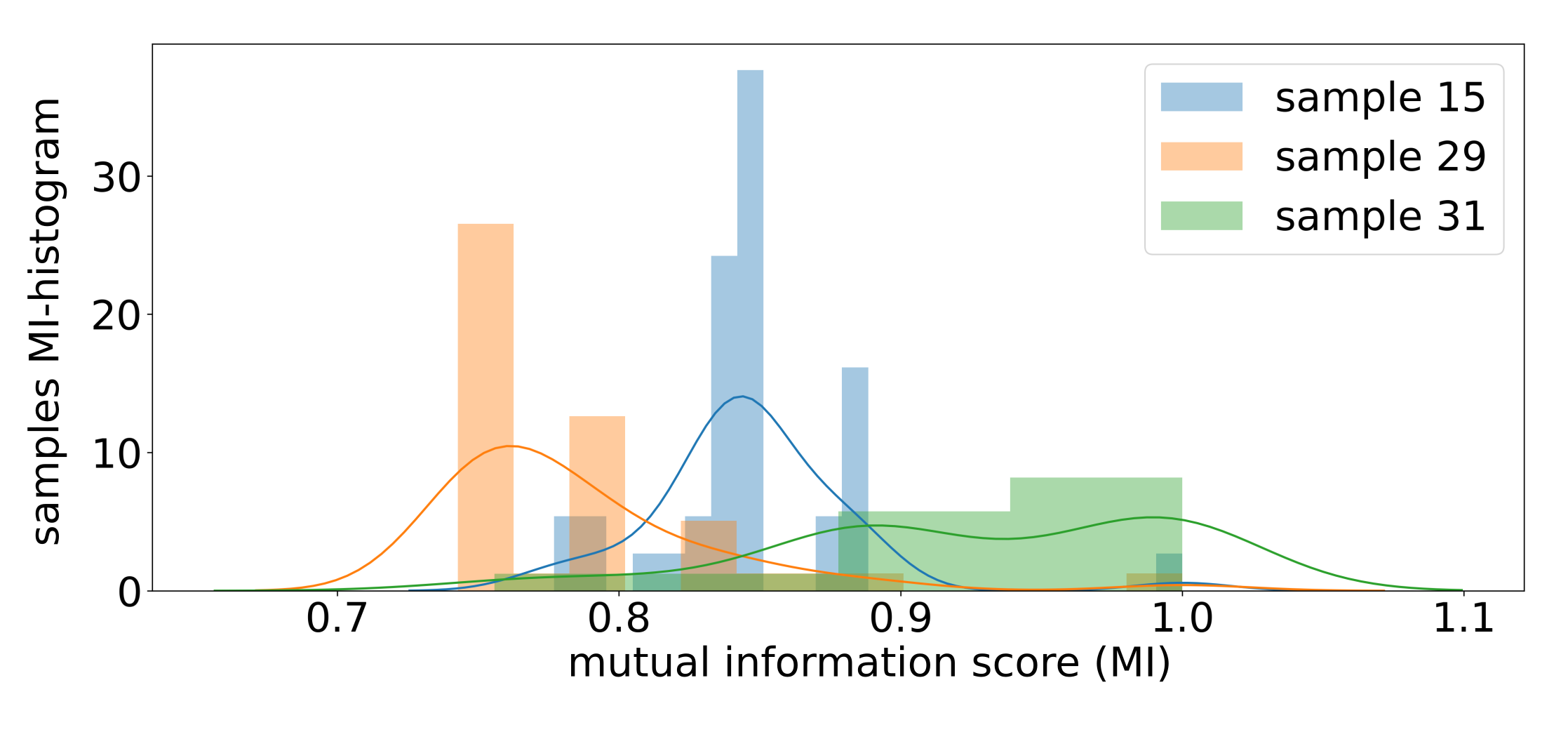}
  \includegraphics[width=0.49\textwidth, clip=True]{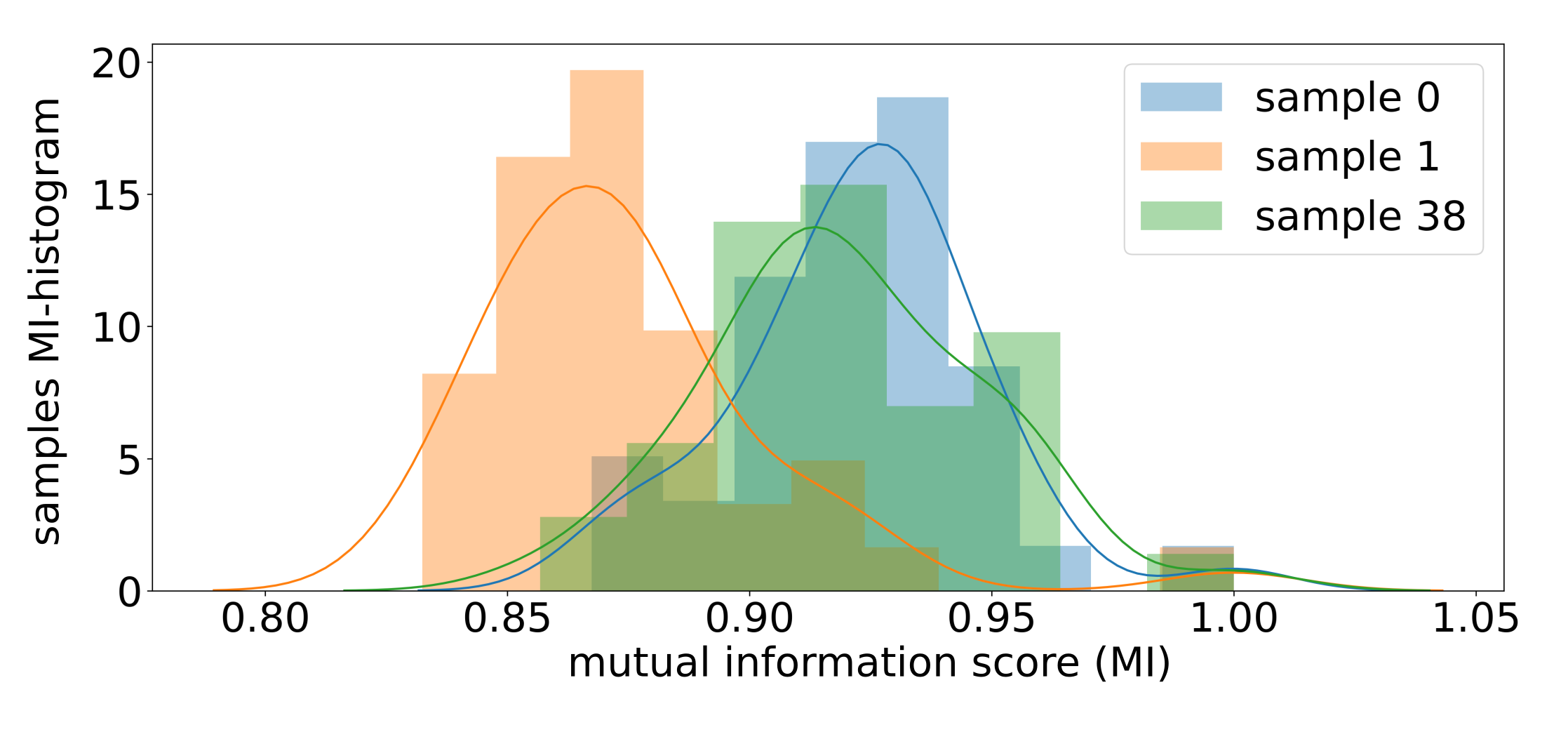}
  \includegraphics[width=0.49\textwidth, clip=True]{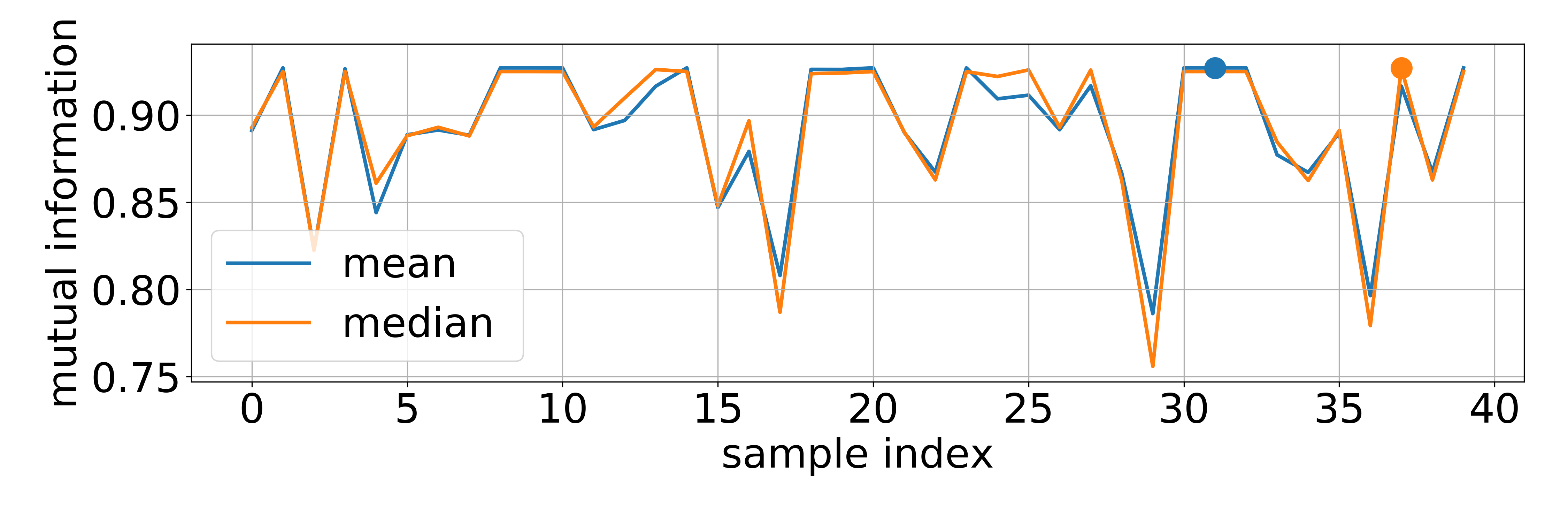}
  \includegraphics[width=0.49\textwidth, clip=True]{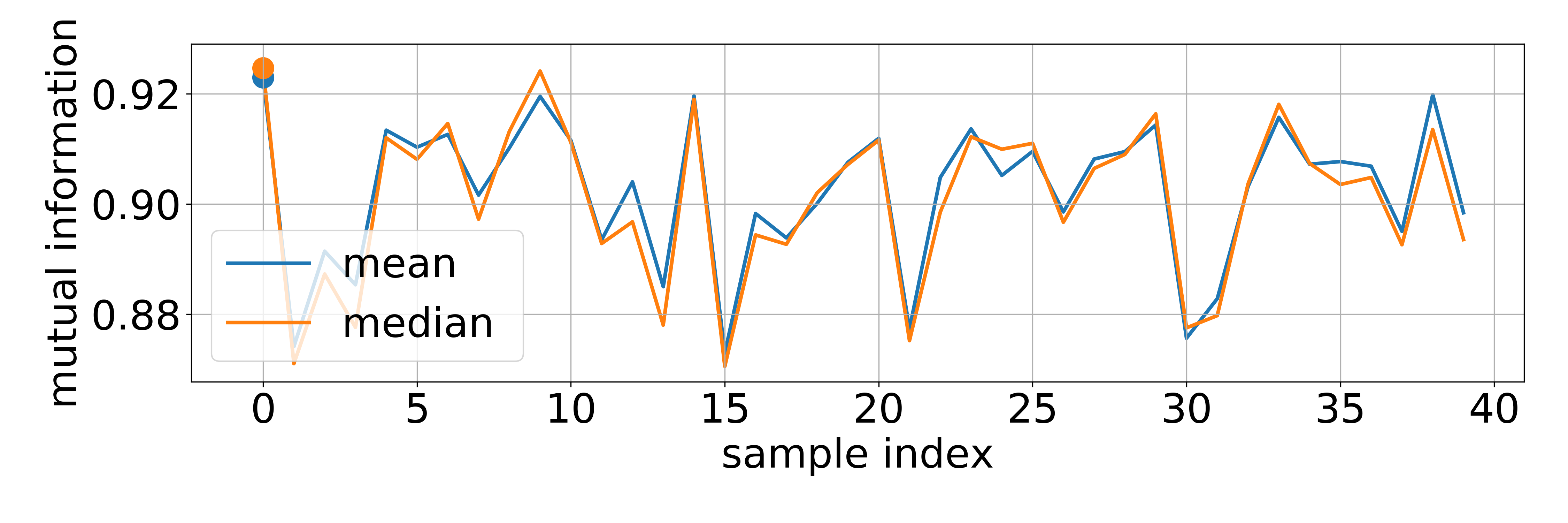}
\caption{Top-left panel: Column-wise histogram of $I_K(\mathbf{k}^*_i,\mathbf{k}^*_j)$, as defined by Eq. (\ref{eq:methods:order:mutual_information}), of clustering samples $\mathbf{k}^*_{i=15}$ (blue), $\mathbf{k}^*_{i=29}$ (orange) and $\mathbf{k}^*_{i=31}$ (green) for the $K=14$ clustering results shown in the left panel of Fig. \ref{fig:wigner:phasediagram:clustering:kmeans:mutual}.    Top-right panel: Column-wise histogram of $I_K(\mathbf{k}^*_i,\mathbf{k}^*_j)$ of clustering samples $\mathbf{k}^*_{i=0}$ (blue), $\mathbf{k}^*_{i=1}$ (orange) and $\mathbf{k}^*_{i=38}$ (green) for the $K=32$ clustering results shown in the right panel of Fig. \ref{fig:wigner:phasediagram:clustering:kmeans:mutual}. Bottom panels: mean (blue) and median (orange) of the adjusted mutual information score, $I_K(\mathbf k^*_i, \mathbf k^*_j)$, for each of the 40 clustering samples, $\mathbf{k}^*_{i}$, with respect to all other 39 clustering samples, $\mathbf{k}^*_{j \ne i}$ shown in Fig. \ref{fig:wigner:phasediagram:clustering:kmeans:mutual} for $K=14$ (left) and $K=32$ (right) clusters (\ie, column-wise average and mean of the data shown in Fig.  \ref{fig:wigner:phasediagram:clustering:kmeans:mutual}). Maxima of the mean and median of the adjusted mutual information score, $I_K(\mathbf k^*_i, \mathbf k^*_j)$, as a function of the 40 sample indices are indicated by filled circles that emphasize clustering results that potentially share the most information with other results on average (or represent the maximum median thereof).}
\label{fig:wigner:phasediagram:clustering:kmeans:mutual:hist}
\end{center}
\end{figure}

For both values of $K$, there is evidence of qualitatively different clustering results to the clustering problem as depicted in Fig. \ref{fig:wigner:phasediagram:clustering:kmeans:mutual}. In order to elucidate this issue we present in Fig.  \ref{fig:wigner:phasediagram:clustering:kmeans:mutual:hist} histograms of the adjusted mutual information score, $I_K(\mathbf{k}^*_i,\mathbf{k}^*_j)$ (as depicted in Fig.  \ref{fig:wigner:phasediagram:clustering:kmeans:mutual} for selected clustering samples), some of which share little information with other clustering results (dark regions in Fig. \ref{fig:wigner:phasediagram:clustering:kmeans:mutual} and orange distributions in Fig. \ref{fig:wigner:phasediagram:clustering:kmeans:mutual:hist}) and others with a more consistent clustering result (bright regions in Fig.  \ref{fig:wigner:phasediagram:clustering:kmeans:mutual} and green distributions in Fig. \ref{fig:wigner:phasediagram:clustering:kmeans:mutual:hist}).

Further, and in order to compare the quality of different clustering results we present in the bottom panels of Fig. \ref{fig:wigner:phasediagram:clustering:kmeans:mutual:hist} the column-wise average value, \ie, $\langle I_K(\mathbf{k}^*_i,\mathbf{k}^*_j)\rangle_j=\sum_{j=0}^{N_c}I_K(\mathbf{k}^*_i,\mathbf{k}^*_j)/N_c$, and the median (analogously defined) of the adjusted mutual information score of each clustering sample, $\mathbf{k}^*_i$, with all other clusterings, $\mathbf{k}^*_{j \ne i}$, and mark both the maximum of the mean and the median. For $K=14$ the clustering sample $i=31$ seems to be a good choice for the final clustering result (cf. left panels of Figs.  \ref{fig:wigner:phasediagram:clustering:kmeans:mutual} and \ref{fig:wigner:phasediagram:clustering:kmeans:mutual:hist}). However, for a larger number of clusters, e.g. $K=32$, it is harder to decide what the optimal clustering might be: both samples (\ie~ for $i=0$ and $i=38$, appear to have qualitatively similar traits as can be seen in the right panels of Figs. \ref{fig:wigner:phasediagram:clustering:kmeans:mutual} and \ref{fig:wigner:phasediagram:clustering:kmeans:mutual:hist}.

However, there is another ingredient that we can include in our analysis to bias the adjusted mutual information score into a physically motivated direction, namely 
the ground state solutions of the {\it symmetric} Wigner bilayer system which also shows up in the phase-diagram of the asymmetric Wigner bilayer system at $A=1$. We can identify the fraction of the data points in the data sets $\mathbf{X}^\mathrm{(asym)}$ which correspond to the ground state solutions of the symmetric Wigner bilayer system and collect them in a separate data set $\mathbf{X}^\mathrm{(sym)}$. We assign all data points in $\mathbf{X}^\mathrm{(sym)}$ to the phases \I~ through \Vs~ following the Table \ref{tbl:labels} \cite{Samaj2012, Samaj2012a} and collect the corresponding phase labels in the set $\mathbf{w}^\mathrm{(sym)}$. Analogously, we collect in the set $\mathbf{k}_i^\mathrm{(sym)}$ the particular clustering labels from the clustering result $\mathbf{k}_i$ (performed on the full data set $\mathbf{X}^\mathrm{(asym)}$ after PCA) which correspond to the data points in $\mathbf{X}^\mathrm{(sym)}$. Hence, the adjusted mutual information score $I_K(\mathbf{w}^\mathrm{(sym)}, \mathbf k_i^\mathrm{(sym)})$ quantifies the overlap between the clustering result, $\mathbf k_i^\mathrm{(sym)}$, and the analytically known labeling, $\mathbf{w}^\mathrm{(sym)}$ (\ie, the amount of commonly labeled data points), of the data set, $\mathbf{X}^\mathrm{(sym)}$, of the feature vectors of the ground states of the symmetric case. We now define the \textit{biased adjusted mutual information score}, $S(\mathbf k_i, \mathbf k_j|\mathbf{w}^{\mathrm{(sym)}})$, via

\begin{equation}
    S(\mathbf k_i, \mathbf k_j|\mathbf{w}^{\mathrm{(sym)}}) =
      I_K(\mathbf k_i, \mathbf k_j) 
      \times
    \sqrt{I_K(\mathbf w^{\mathrm{(sym)}}, \mathbf k_i^{\mathrm{(sym)}})}
    \sqrt{I_K(\mathbf w^{\mathrm{(sym)}}, \mathbf k_j^{\mathrm{(sym)}})},
    \label{eq:wigner:score:biased}
\end{equation}
which weighs the adjusted mutual information, $I_K(\mathbf k_i, \mathbf k_j)$, of different $k$-means (or analogously $k^*$-means \footnote{Also for the $k^*$-means clustering we rely on the set of analytically labeled data, $\mathbf{w}^\mathrm{(sym)}$, of the entire data set, $\mathbf{X}^\mathrm{(asym)}$, which correspond to the ground state solutions of the symmetric case, $A=1$, in the evaluation of $S(\mathbf k^*_i, \mathbf k^*_j|\mathbf{w}^{\mathrm{(sym)}})$, given by \ref{eq:wigner:score:biased}: we respectively compare in $S(\mathbf k^*_i, \mathbf k^*_j|\mathbf{w}^{\mathrm{(sym)}})$ the labels $\mathbf{w}^\mathrm{(sym)}$ with $\mathbf{k}_i^\mathrm{(sym)}$ and $\mathbf{k}_j^\mathrm{(sym)}$, \ie, the fraction of the samples $\mathbf{k}_i^\mathrm{*}$ and $\mathbf{k}_j^\mathrm{*}$ which respectively corresponds to the known ground state structures of the symmetric Wigner bilayer system.}) clustering results, $\mathbf k_i$ and $\mathbf k_j$, with the square root of the respective adjusted mutual information scores of $\mathbf k_i^\mathrm{(sym)}$ and $\mathbf k_j^\mathrm{(sym)}$ with $\mathbf{w}^\mathrm{(sym)}$.

\begin{figure}[ht]
\begin{center}
  \includegraphics[width=0.49\textwidth, clip=True]{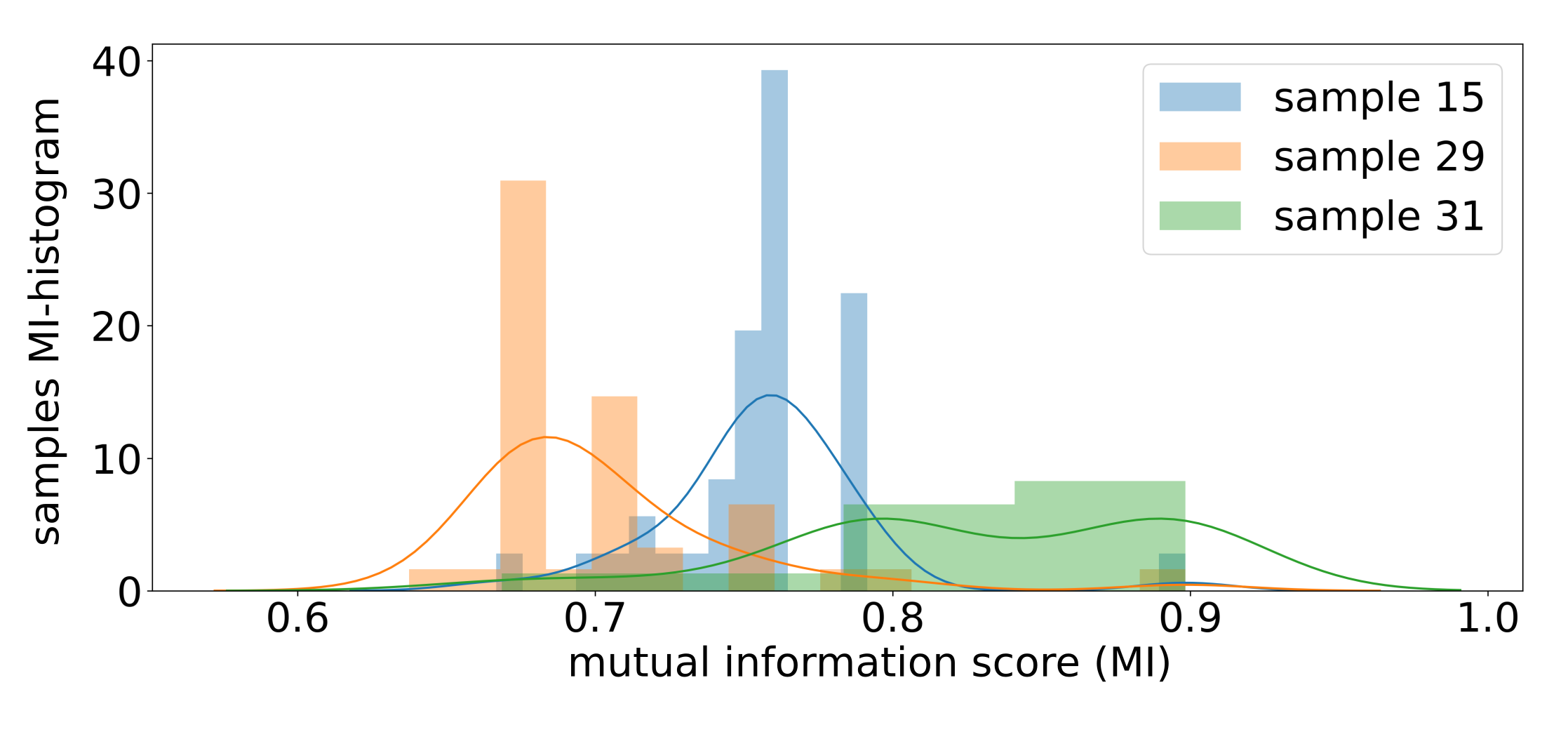}
  \includegraphics[width=0.49\textwidth, clip=True]{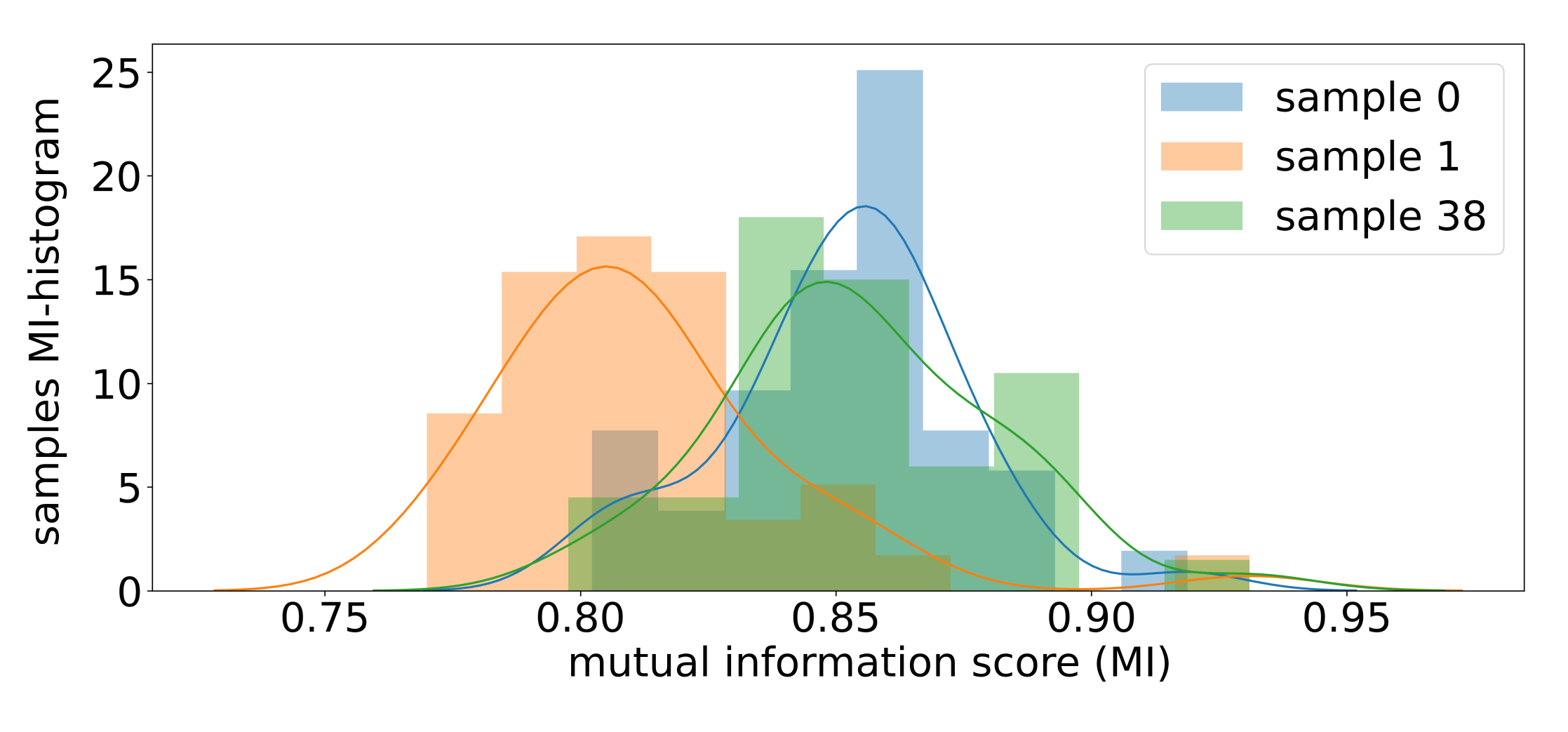}
  \includegraphics[width=0.49\textwidth, clip=True]{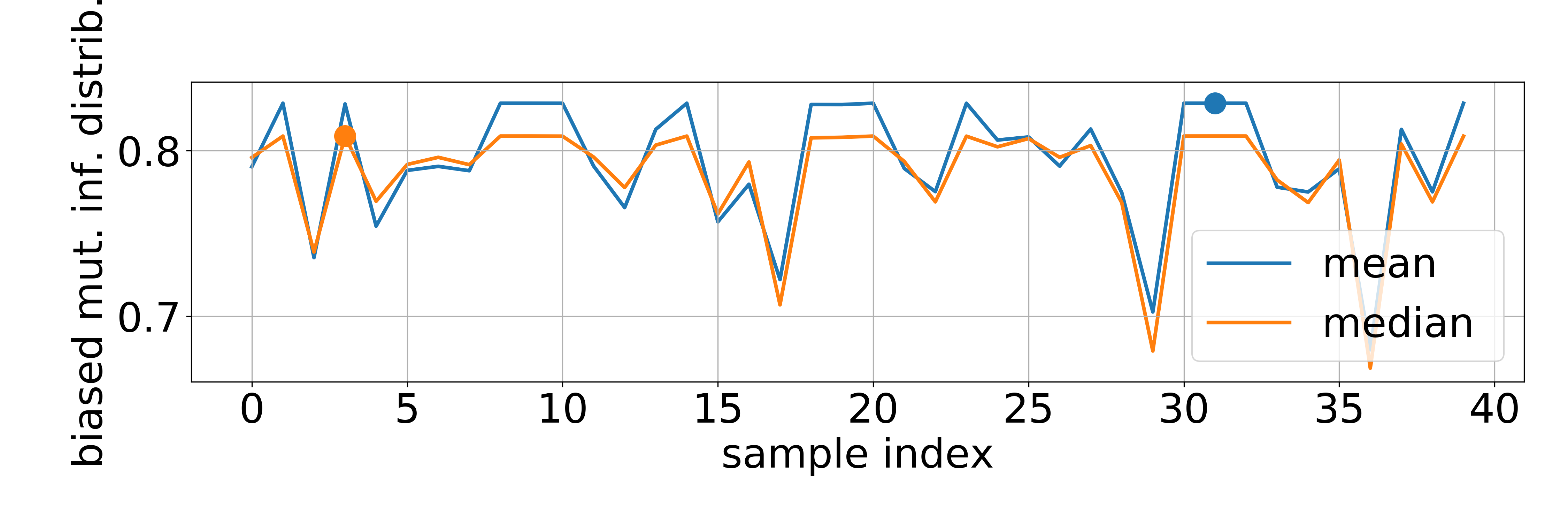}
  \includegraphics[width=0.49\textwidth, clip=True]{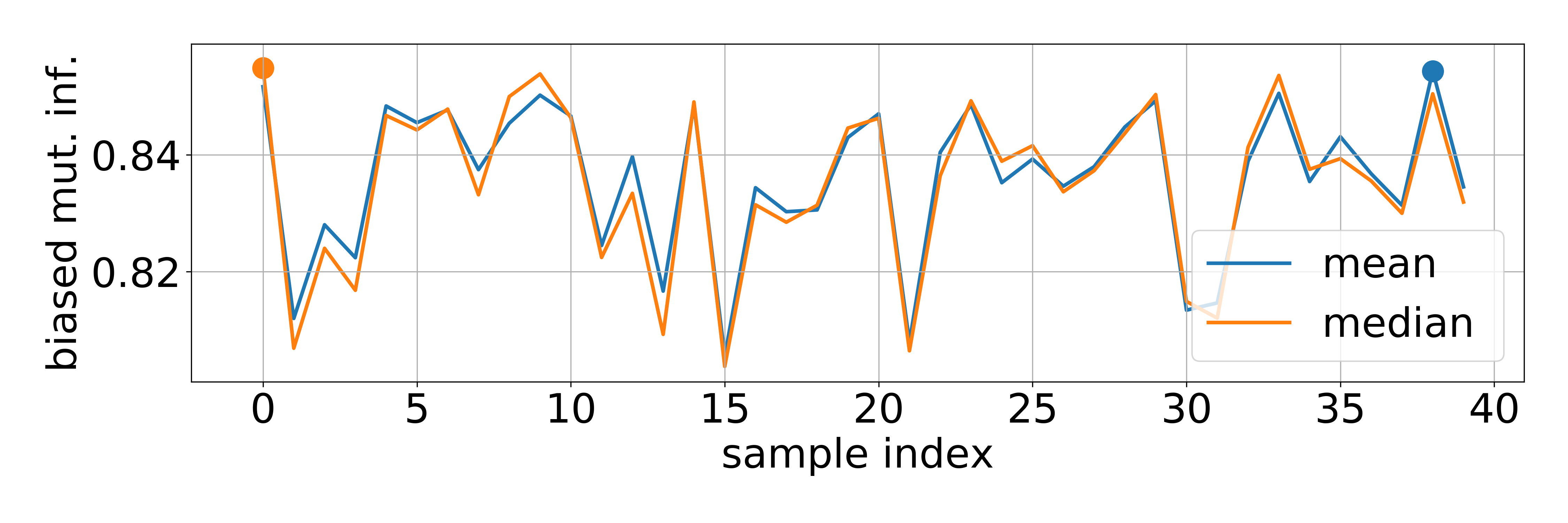}
\caption{Top row and bottom row: same as Fig. \ref{fig:wigner:phasediagram:clustering:kmeans:mutual:hist}
now for the {\it biased} adjusted mutual information score, $S(\mathbf k^*_i, \mathbf k^*_j|\mathbf{w}^{\mathrm{(sym)}})$, defined in Eq. \ref{eq:wigner:score:biased}.}
\label{fig:wigner:phasediagram:clustering:kmeans:mutual:biased}
\end{center}
\end{figure}

In our case, the biased adjusted mutual information score, $S(\mathbf k_i, \mathbf k_j|\mathbf{w}^{\mathrm{(sym)}})$, is an important measure for the quality of the clustering results $\mathbf k_i$ and $\mathbf k_j$ since we demand of a corresponding labeling to be as accurate as possible, especially for the fraction of the data, $\mathbf X^{\mathrm{(sym)}}$, that can be labeled analytically via $\mathbf w^{\mathrm{(sym)}}$. In Fig.~\ref{fig:wigner:phasediagram:clustering:kmeans:mutual:biased} we present the {\it biased} adjusted mutual information score, $S(\mathbf{k}^*_i, \mathbf{k}^*_j|\mathbf{w}^{\mathrm{(sym)}})$, of the same selected samples as used in Fig.~\ref{fig:wigner:phasediagram:clustering:kmeans:mutual:hist} and we also present the corresponding mean and median values of all biased sample scores (\ie, $S(\mathbf k_i, \mathbf k_j|\mathbf{w}^{\mathrm{(sym)}})$) as we have already shown for the unbiased case (\ie, $I_K(\mathbf k^*_i, \mathbf k^*_j)$) in the bottom panels of Fig.~\ref{fig:wigner:phasediagram:clustering:kmeans:mutual:hist}.

By comparing the biased, $S(\mathbf{k}^*_i, \mathbf{k}^*_j|\mathbf{w}^{\mathrm{(sym)}})$, and the unbiased scores, $I_K(\mathbf k^*_i, \mathbf k^*_j)$, we see that in general scaling the adjusted mutual information according to Eq. \ref{eq:wigner:score:biased} leads to smaller values of  $S(\mathbf k^*_i, \mathbf k^*_j|\mathbf{w}^{\mathrm{(sym)}})$ score as compared to $I_K(\mathbf k^*_i, \mathbf k^*_j)$. Especially the diagonal terms, $S(\mathbf k^*_i, \mathbf k^*_i|\mathbf{w}^{\mathrm{(sym)}})$, whose adjusted mutual information scores are $I_K(\mathbf k^*_i, \mathbf k^*_i)=1$ by definition (cf. Fig. \ref{fig:wigner:phasediagram:clustering:kmeans:mutual}), are now weighed by ${I_K(\mathbf{w}^{\mathrm{(sym)}},\mathbf{k}_i^\mathrm{(sym)})\leq1}$, accounting for the quality of the clustering result with respect to the labels of the ground states of the symmetric case. Consequently, the scaling of the adjusted mutual information score, $I(\mathbf{k}^*_i, \mathbf{k}^*_j)$, via Eq. \ref{eq:wigner:score:biased} also causes an additional bias to larger values of the $S(\mathbf k^*_i, \mathbf k^*_j|\mathbf{w}^{\mathrm{(sym)}})$ score for clustering results with large respective overlaps between $\mathbf{k}^{(sym)}_i, \mathbf{k}^\mathrm{(sym)}_j$ and $\mathbf{w}^\mathrm{(sym)}$ (\ie, commonly labeled ground states of the symmetric case);
results with corresponding smaller overlaps of $\mathbf{k}^{\mathrm{(sym)}}_i, \mathbf{k}^\mathrm{(sym)}_j$, and $\mathbf{w}^\mathrm{(sym)}$ are biased towards smaller values of $S(\mathbf{k}^*_i, \mathbf{k}^*_j|\mathbf{w}^{\mathrm{(sym)}})$ (cf. rightmost bins of sample 0 and sample 38 in the top right panel of Figs. \ref{fig:wigner:phasediagram:clustering:kmeans:mutual:hist} and \ref{fig:wigner:phasediagram:clustering:kmeans:mutual:biased}).

We now assume that ``good'' clustering results, which are biased towards large values of the $S(\mathbf k_i, \mathbf k_i|\mathbf{w}^{\mathrm{(sym)}})$ score by labeling the symmetric part in the data set as well as possible, occur frequently and perform similarly in terms of the overall quality of the clustering.
For such good clusterings also the mean (and the median) of the $S(\mathbf k_i, \mathbf k_i|\mathbf{w}^{\mathrm{(sym)}})$ scores are biased towards larger values while being biased towards smaller values for qualitatively poor clustering results.
We define the mean value, $\bar k_i$, of the biased adjusted mutual information score, $S(\mathbf k_i, \mathbf k_i|\mathbf{w}^{\mathrm{(sym)}})$, of the $i,j=0,\dotsc,N_c-1$ different clustering samples (cf. Fig. \ref{fig:wigner:phasediagram:clustering:kmeans:mutual}), by

\begin{equation}
  \bar k_i=\frac{1}{N_c}\sum_{j=0}^{N_c-1}S(\mathbf k_i, \mathbf k_j|\mathbf w^{\mathrm{(sym)}}).
  \label{eq:wigner:score:average}
\end{equation}

With $\bar k_i$ we have a reasonably good measure for comparing different clustering results for one given number of clusters, $K$:
we here rely on $\bar k_i$ to quantify the quality of a clustering result, $\mathbf k^*_i$, of assigning the total number of $K$ clusters correctly, given $N_c$ independent clustering results (cf. Fig. \ref{fig:wigner:phasediagram:clustering:kmeans:mutual:hist}).
We evaluate $\bar k_i$ separately for all independent $k$-means and $k^*$-means clusterings for several different values of $K=14$ to $K=43$:
for a given value of $K$ the one sample from the respective $i=0,\dotsc,N_c-1$ clusterings with the maximum value of $\bar k_i$, given by Eq. \ref{eq:wigner:score:average}, is considered as the best clustering results.



\end{appendix}


\end{document}